\documentclass[twocolumn,english,aps,superscriptaddress,prb,floatfix,longbibliography]{revtex4-2}
\usepackage[utf8]{inputenc} 
\usepackage{color}
\usepackage[dvipsnames]{xcolor}
\usepackage{babel}
\usepackage{amsmath}
\usepackage{amssymb}
\usepackage{graphicx}
\usepackage{esint}
\usepackage{mathtools}

\usepackage{scalefnt}
\usepackage{dcolumn}
\usepackage{bm}
\usepackage{multirow}
\usepackage{wrapfig}
\usepackage[shortlabels]{enumitem}
\usepackage{amsfonts}
\usepackage{physics}
\usepackage{braket}
\usepackage[makeroom]{cancel}
\usepackage{mathrsfs}
\usepackage{ gensymb }
\usepackage{dsfont}
\usepackage{amssymb}
\usepackage[title]{appendix}
\usepackage[dvipsnames]{xcolor}
\usepackage{yfonts}
\usepackage[export]{adjustbox}
\usepackage{pagecolor}
\usepackage{tcolorbox}
\usepackage{diagbox}
\usepackage{booktabs}

\usepackage[unicode=true,pdfusetitle,
bookmarks=true,bookmarksnumbered=false,bookmarksopen=false,
breaklinks=true,pdfborder={0 0 0},pdfborderstyle={},backref=false,colorlinks=true]
{hyperref}
\hypersetup{
linkcolor=blue,citecolor=blue,urlcolor=blue}

\newcommand{\pa}[1]{\left(#1\right)}
\newcommand{\co}[1]{\left[ #1\right] }

\newcommand{\mean}[1]{\left\langle #1 \right\rangle }

\newcommand{\q}{\textbf{q}}

\newcommand{\gb}{\overline{g}}

\newcommand{\br}{\textbf{r}}

\newcommand{\betab}{\boldsymbol{\beta}}

\newcommand{\cb}{\boldsymbol{c}}
\newcommand{\wbb}{W}
\newcommand{\xbb}{X}
\newcommand{\ybb}{Y}

\newcommand{\lamb}[1]{\lambda^\mu_{\textbf{q},#1}}
\newcommand{\Lamb}[1]{\Lambda^\mu_{\textbf{q},#1}}
\newcommand{\tLamb}[1]{\widetilde{\Lambda}^{\mu}_{\textbf{q},#1}}
\newcommand{\tLambij}[2]{\widetilde{\Lambda}^{#1}_{\textbf{q},#2}}
\newcommand{\tLambmij}[2]{\widetilde{\Lambda}^{#1}_{-\textbf{q},#2}}
\newcommand{\tL}[2]{\widetilde{\Lambda}^{#1}_{#2}}

\newcommand{\rucl}{$\alpha$-RuCl$_3$}
\newcommand{\hlio}{H$_3$LiIr$_2$O$_6$}

\makeatother

\begin{document}
\title{Phonon dynamics in the site-disordered Kitaev spin liquid }

\author{Vitor Dantas}
\affiliation{School of Physics and Astronomy, University of Minnesota, Minneapolis, MN 55455, USA}
\author{Wen-Han Kao}
\affiliation{School of Physics and Astronomy, University of Minnesota, Minneapolis, MN 55455, USA}
\author{Natalia B. Perkins}
\affiliation{School of Physics and Astronomy, University of Minnesota, Minneapolis, MN 55455, USA}

\begin{abstract}
	The Kitaev honeycomb model provides a paradigmatic example of an exactly solvable quantum spin liquid
	(QSL), in which the spin degrees of freedom fractionalize into itinerant Majorana fermions coupled to a static background of $\mathbb{Z}_{2}$ gauge fluxes. This model has attracted significant attention in recent years due to the possibility of its experimental realization in some spin-orbit Mott insulators such as $ \alpha $-RuCl$ _3 $. Among various experimental probes, ultrasound experiments measuring sound attenuation have emerged as a promising avenue to unveil the fractionalization of spins in these materials. Yet, candidate materials often deviate from the ideal Kitaev model due to the presence of disorder, leading to the emergence of localized modes governing low-energy physics. To provide further insight into the effects of these defect-induced modes on the phonon dynamics, we calculate the sound attenuation coefficient in the site-disordered Kitaev honeycomb model with an applied magnetic field, which breaks the time-reversal symmetry. In order to obtain a more accurate perspective on the temperature-dependent sound attenuation in this model, the impact of thermally excited fluxes on the disordered system is also analyzed.
	
\end{abstract}
\date{\today}

\maketitle

\section{Introduction}

In recent decades, significant efforts have been directed toward discovering and synthesizing compounds that manifest quantum spin-liquid (QSL) phases. These enigmatic states of matter exhibit exotic phenomena such as long-range entanglement, emergent gauge theories, and fractionalized spin excitations. Central to this pursuit are Kitaev materials ~\cite{jackeli09,chaloupka10,chaloupka13,winter17,trebst17,takagi19,Rousochatzakis2024}, which feature dominant Ising-like bond-dependent interactions for $j_{{\text{eff}}}=1/2$  effective moments on a honeycomb lattice and hold promise for realizing Kitaev QSL
~\cite{kitaev06}. 
This model admits an exact solution, revealing a quantum spin-liquid phase with fractionalized excitation represented by gapless/gapped Majorana fermions and static $\mathbb{Z}_{2}$ fluxes ~\cite{kitaev06,baskaran07,savary17b}.

At the forefront of this rapidly growing field is the search for signatures of spin fractionalization in candidate materials, such as the honeycomb iridates $A_{2}$IrO$_{3}$ ($A=\mathrm{Li,Na}$)~\cite{choi12,singh12} and \rucl ~\cite{plumb14,sears15,banerjee2016proximate}. Various dynamical probes are employed in this search, including inelastic neutron scattering ~\cite{knolle14,knolle15,banerjee2016proximate}, Raman scattering ~\cite{knolle2014raman,perreault2015theory,perreault2016resonant,rousochatzakis2019quantum,sahasrabudhe2020high,wang2020range,YanyYang2021,yang2022signatures}, resonant inelastic X-ray scattering ~\cite{Gabor2016,halasz2017probing,halasz2019observing,ruiz2021magnon}, ultrafast spectroscopy \cite{alpichshev2015confinement}, terahertz nonlinear coherent spectroscopy ~\cite{wan2019resolving,little2017antiferromagnetic}, and inelastic scanning tunneling microscopy (STM) ~\cite{pereira2020electrical,udagawa2021scanning,joy2022dynamics,bauer2023scanning,kao2024dynamics,kao2024STM,Bauer2024STM}.
 
 Among these several routes, recent studies have highlighted phonon dynamics as an insightful probe for understanding quantum spin liquids in candidate materials, given that spin-lattice coupling is inevitable and often strong in materials with large spin-orbit coupling \cite{kasahara18a,li2021giant}.
  For instance, sound attenuation has been  shown to be 
  a useful tool to provide information about two-dimensional frustrated magnetic systems, including quantum spin liquids \cite{Lee2011PhysRevLett.106.056402,serbyn2013spinon,metavitsiadis2020phonon,ye2020phonon,simon2022ultrasound,Hauspurg2024,singh2023phonon}. In the context of Kitaev materials, previous investigations have shown that the Majorana fermion-phonon coupling gives rise to unique signatures such as a linear temperature-dependent sound attenuation and a characteristic sixfold angular dependence from the lattice symmetry \cite{ye2020phonon,metavitsiadis2020phonon,feng2021temperature}. Other works also explored the effects of magnetoelastic coupling between spins and optical phonons, which may be accessible via Raman spectroscopy \cite{perreault2015theory,perreault2016resonant,yang2022signatures,feng2022footprints,metavitsiadis2022optical}. 

 Here we propose that the phonon dynamics is a meaningful experimental probe of the Kitaev materials, even with inevitable disorder.
  The role of disorder in these systems  gained a lot of attention in recent years, especially after the synthesis of the iridate compound 
  \hlio, known for its unique thermodynamic behavior  \cite{kitagawa18,lee2023coexistence,de2023momentum}. 
Disorder is also crucial in other candidate materials like Ir-doped RuCl$_{3}$, where vacancies play a major role \cite{baek20}. Extensive theoretical effort has been made to understand disorder in the Kitaev spin liquid, with minimal models incorporating bond randomness and various types of vacancies ~\cite{willans10,zschocke15,knolle19,nasu20,kao2021localization,kao2021vacancy,nasu21,dantas2022}. It was shown that disorder dramatically changes the low-energy excitations in the Kitaev model, impacting dynamical and thermodynamic properties such as spin dynamics, specific heat, and susceptibility ~\cite{knolle19,kao2021vacancy,nasu21,dantas2022,kao2024dynamics,kao2024STM}, and even topological properties in the non-Abelian regime ~\cite{yamada20,dantas2022,Vlad2024}.
To achieve an effective and realistic description of dynamical probes in candidate materials, it is essential to account for these defects. 

Therefore, in this paper, we aim to extend our previous investigations of phonon dynamics in the Kitaev model by incorporating disorder in the form of quasivacancies ~\cite{kao2021vacancy}. Specifically, we compute the sound attenuation for an effective spin-lattice coupled Kitaev spin liquid with quasivacancies. As sound attenuation originates from dissipative scattering processes, it can be connected to the imaginary part of the phonon polarization bubble using the perturbative methods developed in Ref.~\cite{ye2020phonon} and the real-space formalism introduced in Ref.~\cite{feng2021temperature} to incorporate site disorder.

Our analysis demonstrates how the intricate low-energy physics of quasilocalized modes \cite{kao2021vacancy,dantas2022,kao2024dynamics} in the presence of quasivacancies is reflected in the phonon self-energy. We show that the sound attenuation coefficient exhibits the expected sixfold symmetry, as in the clean system \cite{ye2020phonon}. Crucially, the sound attenuation remains linear at low temperatures even in the presence of disorder, indicating the robustness of this observable in probing fractionalization. Moreover, we show that this behavior may persist in the presence of an external field due to the scattering of quasilocalized modes induced by quasivacancies into the bulk. This is in stark contrast to the pure-model description, where the gapped Majoranas give a negligible contribution to sound attenuation at low temperatures ~\cite{ye2020phonon}.
Finally, we demonstrate that the attenuation coefficient continues to grow linearly with temperature in the random-flux regime. This behavior is fundamentally distinct from other possible physical processes, reinforcing that fractionalized degrees of freedom dominate the attenuation process even in the presence of disorder.

The rest of the paper is organized as follows: In Sec.\ref{Sec_model}, we introduce the model of this work. The disordered Kitaev model, the phonon Hamiltonian, and the spin-phonon coupling are described in Sec. \ref{sec_spinham}, \ref{sec_phononham} and \ref{sec_spin-phham}, respectively. Next, we show the computation of the polarization bubble in the mixed representation in Sec. \ref{sec_phonon_dyn}. In Sec. \ref{sec_soundatt}, we present our results for the sound attenuation, where a simple argument for the temperature dependence is shown in Sec. \ref{sec_powercount}. We present the numerical results for quasivacancies and flux disorder in Secs. \ref{sec_soundatt_Quasivac} and  \ref{sec_soundatt_RF}, respectively. We close with a summary in Sec.\ref{conclusions}. Additional details and results are presented in Appendices \ref{App: calc_details} and \ref{App:bound-flux}.

%
%

\section{The Model}\label{Sec_model}

In this section, we introduce the model and review the site-disordered Kitaev spin liquid and the derivation of the spin-phonon interaction in the presence of disorder.

 The full Hamiltonian comprises the  pure spin and phonon terms, along with the spin-phonon interaction:
\begin{align}
	\mathcal{H} = \mathcal{H}_{\text{spin}} + \mathcal{H}_{\text{ph}} + \mathcal{H}_{\text{spin-ph}}
	\label{eq:fullH}
\end{align}
The first term is the Kitaev model in the presence of a time-reversal symmetry-breaking field~\cite{kitaev06}, written in real space to incorporate disorder. The second term is the two-dimensional free-phonon Hamiltonian on the honeycomb lattice in the long-wavelength limit, focusing on low-energy acoustic phonon modes. We assume disorder affects only the magnetic interactions, not the lattice structure, so $\mathcal{H}_{\text{ph}}$ is represented in momentum space. The last term describes the magnetoelastic coupling, written in a mixed representation.
  

\subsection{ Spin Hamiltonian}\label{sec_spinham}

\begin{figure*}
\centering
\includegraphics[width=0.9\linewidth]{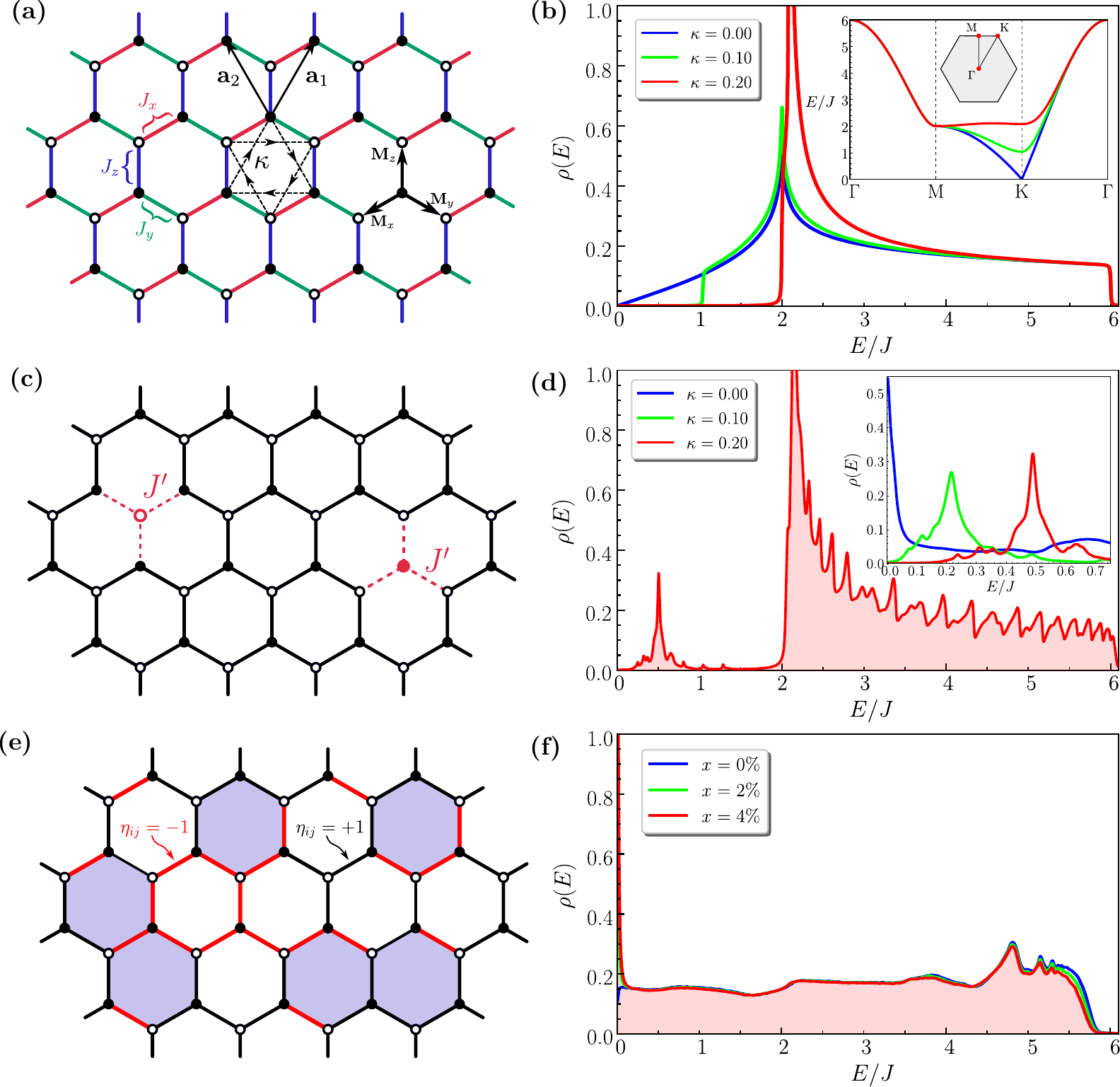}
\caption{(a)  Kitaev honeycomb model: The three bond-dependent Kitaev exchanges $J_\alpha$ are shown  by red, green and blue bonds. The arrows  $\textbf{a}_1 = (1,\sqrt{3})/2$ and $\textbf{a}_2 = (-1,\sqrt{3})/2$ denote the lattice vectors, and $\textbf{M}_x, \textbf{M}_y$, and $\textbf{M}_z$ are the  nearest-neighbor vectors. 
 At the central plaquette, the next nearest-neighbor effective interaction $\kappa$ is shown.
(b) Density of states (DOS) of the clean model for different values of $\kappa$. Inset: Majorana-fermion dispersion. The $\kappa$ interaction opens a gap at the Dirac point ($\textbf{K}$-point).
(c) Quasivacancies in the zero-flux sector. The couplings of a quasivacancy to the rest of the lattice  $J^\prime$ are shown by dashed red lines.
(d) DOS in the zero-flux sector with quasivacancies for $\kappa = 0.2$. Inset: In-gap peak for different $\kappa$ values. Here we set $x = 2\%$ and $J^\prime = 0.01 J$.
(e) Random-flux sector without quasivacancies: Negative (positive)  link variables are shown by red (black) links. Shaded hexagons represent the $\pi$-fluxes.
(f) DOS of the random-flux sector for different concentrations of quasivacancies with $\kappa = 0$. The low-energy peak indicates localized modes due to finite $x$. Away from the low-energy limit, the DOS is approximately flat for all concentrations.  In disordered systems, plots (d) and (f), 1000 realizations with $L=32$ are calculated and averaged.}
\label{fig:fig1}
\end{figure*}

The spin-1/2  Kitaev Hamiltonian on a honeycomb lattice in the presence of a time-reversal symmetry-breaking field reads \cite{kitaev06}:
\begin{align}\label{kitaev_def}
  \mathcal{H}_{\text{spin}} = -\sum_{\left \langle ij \right \rangle}J_{\left \langle ij \right \rangle_{\alpha}}{\sigma}_{i}^{\alpha}{\sigma}_{j}^{\alpha} - \kappa\sum_{\left \langle \left \langle ik \right \rangle \right \rangle_{\alpha,\beta}}{\sigma}_{i}^{\alpha}{\sigma}_{j}^{\beta}{\sigma}_{k}^{\gamma},
\end{align} 
where ${\sigma}^{\alpha}_{i}$ denotes Pauli spin operators with $\alpha = x, y, z$ and $\left \langle ij \right \rangle_{\alpha}$ labels the nearest-neighbor sites $i$ and $j$ along an $\alpha$-type bond.   The second term is the three-spin interaction with strength $\kappa \sim \frac{h_{x}h_{y}h_{z}}{J^{2}}$  that imitates an external magnetic field and breaks time-reversal symmetry while preserving the exact solvability ~\cite{kitaev06}. By rewriting each spin operator in terms of four Majorana fermions, ${\sigma}^{\alpha}_{i} = i{b}^{\alpha}_{i}{c}_{i}$, and defining the link operators ${\eta}_{ij}^{\alpha}=i{b}^{\alpha}_{i}{b}^{\alpha}_{j}$, the Hamiltonian takes the form
\begin{align} \label{eq:kitaev_majoranas}
  \mathcal{H}_{\text{spin}} = i\sum_{\left \langle ij \right \rangle}J_{\left \langle ij \right \rangle_{\alpha}}{\eta}_{ij}^{\alpha}{c}_{i}{c}_{j} + i\kappa\sum_{\left \langle \left \langle ik \right \rangle \right \rangle_{\alpha,\beta}}{\eta}_{ij}^{\alpha}{\eta}_{kj}^{\beta}{c}_{i}{c}_{k}.
\end{align}

In the following, we restrict ourselves within the isotropic limit of the pristine model, $J_\alpha \equiv J$, and 
focus predominantly on a kind of disorder dubbed quasivacancy \cite{kao2021vacancy,kao2021localization}. Contrary to a true vacancy, where a site is completely removed from the lattice, a quasivacancy is defined by a weakly interacting site with respect to its neighbors. More precisely, the Kitaev exchange for a quasivacancy is given by $J^\prime \ll J$. 
 Physically, this can be identified as non-magnetic defects that are weakly connected to their neighbors due to extremely strong but relatively rare bond randomness.
 In order to introduce randomly distributed quasivacancies, we  rewrite the spin Hamiltonian 
 (\ref{eq:kitaev_majoranas}) as \cite{kao2021vacancy}:
 \begin{align}
	 \mathcal{H}_{\text{spin}} =  \,& iJ\sum\limits_{\substack{\left\langle ij\right\rangle _{\alpha} \\ i,j \in \mathcal{P}} }
	\eta_{ij}^{\alpha}c_{i}c_{j} + iJ^\prime\sum\limits_{\substack{\left\langle ij\right\rangle _{\alpha} \\ i \in \mathcal{Q}, j \in \mathcal{P}} }\eta_{ij}^{\alpha}c_{i}c_{j}  \nonumber\\
	&+ i\kappa\sum_{\left\langle \left\langle ik\right\rangle \right\rangle_{\alpha,\beta} }\eta_{ij}^{\alpha}\eta_{kj}^{\beta}c_{i}c_{k},
\end{align}
 Where $\mathcal{P}$ denotes the set of normal sites (the bulk) and $\mathcal{Q}$ is the set of quasivacancies. 

In the pristine model with $\kappa = 0$, the flux configuration that minimizes the energy is the zero-flux sector, that is, $W_{p}=\prod_{\left\langle ij\right\rangle \in p}\eta_{ij}^{\alpha}= +1$, for all $p$.
This results in a straightforward tight-binding model for Majorana fermions, which can be easily diagonalized in momentum space. The low-energy dispersion of the Majorana fermions exhibits a graphene-like behavior, characterized by two Dirac cones at the $\textbf{K}$ and $\textbf{K}^\prime$ points of the Brillouin zone. In the presence of a finite field $(\kappa \neq 0)$, the Majorana excitations become gapped, with an energy gap given by  $\Delta_\kappa/J = \text{max}\pa{6\sqrt{3}\kappa, 2}$,  as shown in Fig.~\ref{fig:fig1}(b).

Introducing quasivacancies disrupts the translation symmetry, preventing diagonalization in reciprocal space and necessitating real-space diagonalization. Despite this, the model remains exactly solvable. Moreover, unlike the clean limit, the ground state of the Kitaev model with quasivacancies does not necessarily correspond to the zero-flux sector. It was shown that
the energy is lowered by binding a flux to the extended plaquette around the vacancy \cite{willans10,Willans2011}.
 The flux-binding effect persists even with a finite concentration of vacancies, including quasivacancies weakly coupled to the rest of the system ($J'\ll J$) \cite{kao2021vacancy}.
 However, as our numerical results have shown that the flux-binding effect does not affect the sound attenuation response (see Appendix \ref{App:bound-flux}), we will focus on the zero-flux and random-flux sectors only.

In a given flux
sector,  the gauge can be fixed by specifying all link variables $\{\eta_{ij}\}$. The resulting Hamiltonian \eqref{eq:kitaev_majoranas} of non-interacting Majorana fermions can be rewritten in the following matrix form:
\begin{align}
	\mathcal{H}_{\text{spin}}& = \frac{i}{2}\label{Hmaj matform}
	\begin{pmatrix}
		\boldsymbol{c}_A & \boldsymbol{c}_B
	\end{pmatrix}
	\begin{pmatrix}
		F & M\\
		-M^T& -D
	\end{pmatrix}
	\begin{pmatrix}
		\boldsymbol{c}_A\\ \boldsymbol{c}_B
	\end{pmatrix}, 
\end{align}
where $\boldsymbol{c}_{A(B)^{}}$ denote the $N-$component vectors with components $c_{i,A(B)}$ for a lattice with $N = L^2$ unit cells. The matrix $M_{ij} = J_{ij} \eta_{ij}$  defines the hopping between different sublattices, with $J_{ij}=J$ if the bond is in the bulk and $J_{ij}=J'$ if the bond involves a quasivacancy ~\cite{kao2021vacancy}. 
The hopping within the same sublattice is represented by the matrices
$F_{ik} = \kappa \eta_{ij}^\alpha \eta_{kj}^\beta$ (on sublattice $A$) and $D_{ik} = \kappa \eta_{ij}^\alpha \eta_{kj} ^\beta$ (on sublattice $B$), which have non-zero elements only when $\kappa\neq 0$.
Note that, generally,  $F$ and $D$ do not necessarily have to be identical due to the sublattice symmetry breaking induced by a generic flux configuration. 
 In our algorithm, we select a randomly distributed percentage $x$ of spins from the system to define the quasivacancy subspace $\mathcal{Q}$, ensuring a balanced distribution of quasivacancies. This means the number of quasivacancies on each sublattice is
 $N_A = N_B = xN/2$.

To bring the Hamiltonian (\ref{Hmaj matform}) into its canonical form, we introduce the complex fermion operators 
$f^\dagger$, which are related to the Majorana fermions by the transformation
 $U$, defined as:
\begin{align}
	\begin{pmatrix}
		f\\ f^\dagger
	\end{pmatrix} = 
	\frac{1}{2}\begin{pmatrix}
		1 & i\\
		1 & -i
	\end{pmatrix}
	\begin{pmatrix}
		\boldsymbol{c}_A\\ \boldsymbol{c}_B
	\end{pmatrix}
	=U^{-1}	\begin{pmatrix}
		\boldsymbol{c}_A\\ \boldsymbol{c}_B
	\end{pmatrix}.
	\label{comptransf}
\end{align}
In this complex fermion basis, the Hamiltonian assumes the Bogoliubov de-Gennes form \cite{blaizot1986,knolle15},
\begin{align}\label{Hcomplex}
	\mathcal{H}_{\text{spin}}  = \frac{1}{2}\;
	\begin{pmatrix}
		f^\dagger & f
	\end{pmatrix}
	\begin{pmatrix}
		h&\Delta\\
		\Delta^\dagger&-h^T
	\end{pmatrix}
	\begin{pmatrix}
		f\\
		f^\dagger
	\end{pmatrix},
\end{align}
with the $N\times N$ matrices $\Delta$ and $h$ defined in terms of $M$, $D$, and $F$ as
\begin{align}
\begin{split}
	&\Delta = (M^T - M) + i(F+D), \\
	&h = (M + M^T) + i(F - D). 
\end{split}
\end{align}
To bring the matrix in Eq.~\eqref{Hcomplex} into its diagonal form, we define the unitary transformation $W$, which is the matrix representation of the Bogoliubov transformation \cite{blaizot1986}:
\begin{align}
	W \begin{pmatrix}
		h&\Delta\\
		\Delta^\dagger&-h^T
	\end{pmatrix}
	W^\dagger = 
	\begin{pmatrix}
		E& 0\\
		0&-E
	\end{pmatrix}.
\end{align}
Here, $E$ is the $N\times N$
matrix containing the positive eigenvalues
 $\varepsilon_i$ stored in descending order. With this convention, it is easy to see that $W^\dagger$ is the matrix with the eigenvectors stored columnwise as $\wbb^\dagger = \pa{\textbf{V}^+_1\, \dots \textbf{V}^+_N \; \textbf{V}^-_1\, \dots \textbf{V}^-_N}$, where $\textbf{V}^{+(-)}_n$ is the eigenvector corresponding to the $n-$th positive (negative) eigenvalue $\pm\varepsilon_n$. The list of positive eigenvalues $(\varepsilon_N,\dots ,\varepsilon_1 )$ and the matrix $\wbb^\dagger$ 
 constitute our numerical output from the exact diagonalization.
 
 Following the notation used in Ref. 
   \cite{blaizot1986}, we introduce $N\times N$ Bogoliubov matrices $X$ and $Y$. Thus, the operator 
 $W$ can be written as:
\begin{align}\label{bogomat}
	\wbb = 
	\begin{pmatrix}
		\xbb^* & \ybb^*\\
		\ybb  & \xbb
	\end{pmatrix} \qquad
	\wbb^\dagger = 
	\begin{pmatrix}
		\xbb^T & \ybb^\dagger\\
		\ybb^T  & \xbb^\dagger
	\end{pmatrix} 
\end{align}
Now, we can define the Bogoliubov quasiparticles  $\beta$ and $\beta^\dagger$, which are related to $f$ and $f^\dagger$ via 
\begin{align}\label{bogtrans}
\begin{split}
&\beta_i = \sum_{j} \left (X^{T}_{ij} f_j^{} + Y_{ij}^\dagger f_j^\dagger\right )\\
&\beta_i^\dagger = \sum_{j} \left (Y^{T}_{ij} f_j^{} + X_{ij}^\dagger f_j^\dagger\right ),
\end{split}
\end{align}

where $X_{ij}$ and $Y_{ij}$ are defined in \eqref{bogomat}. 
In terms of the Bogoliubov quasiparticles,  $\mathcal{H}_\text{spin}$ becomes 
\begin{align}\label{diagform}
	\mathcal{H}_\text{spin} = \sum_i\varepsilon_i \pa{\beta_i^\dagger \beta_i - \frac{1}{2}} . 
\end{align}

Finally, the ground state is defined as the state with no quasiparticle excitations, i.e. $\beta_i\ket{0} = 0$. From this definition, along with \eqref{diagform}, we define the ground-state energy as $E_0 = -\frac{1}{2}\sum_{i}\varepsilon_i$  and the density of states (DOS) as $\rho(\varepsilon) = \sum_{i}\delta(\varepsilon - \varepsilon_i)$. 
As shown in recent works \cite{kao2021vacancy,kao2021localization} the effects of quasivacancies are well pronounced
even at very low concentrations. More specifically, when the field is absent ($\kappa = 0$), there is a pileup of low-energy states, as shown in the inset of Fig.~\ref{fig:fig1}(d). The pileup of low-energy states emerges for any finite $x$, and its appearance is independent of the flux-sector \cite{kao2021localization}.


We also consider the effect of flux disorder, originating from the thermal proliferation of fluxes at finite temperatures.  This intrinsic property of the Kitaev model is reflected in a specific heat crossover in the 2D model at temperatures near the flux-excitation energy scale \cite{nasu15}. 
 In the high-temperature limit, it has an equal probability to have zero- or $\pi$-flux on each plaquette.
A realization of the random-flux sector is shown in Fig.~\ref{fig:fig1}(e). This introduces additional disorder to the Majorana fermions as a random-sign hopping problem.  It significantly affects the density of states, smearing out both the Van Hove singularity and the Dirac cone (see  Fig.~\ref{fig:fig1} (f)), and has substantial impacts on both thermodynamic and dynamical responses \cite{rousochatzakis2019quantum,halasz2019observing,feng2022footprints}.
Previous studies have considered this type of disorder in phonon dynamics within the Kitaev model \cite{metavitsiadis2020phonon, feng2021temperature}. Here, we extend the analysis by extracting the temperature dependence of the sound attenuation coefficient in the presence of both flux and quasivacancy disorders, a topic not previously explored.\\




\subsection{Phonon Hamiltonian}\label{sec_phononham}

Since the type of disorder studied in this work does not affect the lattice symmetry -- quasivacancies only alter the magnetic exchange between an ion and its neighboring sites -- the description of acoustic phonons remains the same for both the clean model and the model with quasivacancies. Then,  following  Refs.~\cite{ye2020phonon,feng2021temperature}, we assume a homogeneous elastic medium in the long wavelength limit to write an effective action in terms of
the strain tensor $\epsilon_{ij} = \frac{1}{2}(\partial_i u_j +\partial_j u_i)$, where $u_i$ are the displacement field components, and the independent components of the elastic modulus tensor 
 $\mathcal{C}_{ijkl}$. We then impose the $C_{6v}$ point group symmetry of the honeycomb lattice to write the elastic energy in the basis of $A_1^{\text{ph}}$ and $E_2^{\text{ph}}$ irreducible representations (IRRs) of the point group.

Decomposing the displacement fields into two independent polarizations, $\textbf{u}_\textbf{q} = \sum_{\mu}e^\mu_\textbf{q}\tilde{\textbf{u}}^\mu_\textbf{q}$, where
$e^\mu_\textbf{q}$ are the polarization vectors for $\mu = \parallel,\perp$ (longitudinal and transverse polarizations), and diagonalizing $\mathcal{H}_{\mathrm{ph}}$,  we obtain the phonon dispersion and the corresponding polarization vectors:
\begin{align}
\begin{split}    
	& \Omega_{\mathbf{q}}^{\|}=v_s^{\|} q=\sqrt{\frac{C_1+C_2}{\rho}} q, \quad \hat{e}_{\mathbf{q}}^{\|}=\left\{\cos \theta_{\mathbf{q}}, \sin \theta_{\mathbf{q}}\right\}, \\
	& \Omega_{\mathbf{q}}^{\perp}=v_s^{\perp} q=\sqrt{\frac{C_2}{\rho}} q, \quad \hat{e}_{\mathbf{q}}^{\perp}=\left\{-\sin \theta_{\mathbf{q}}, \cos \theta_{\mathbf{q}}\right\},
\end{split}
\end{align}
where $\theta_{\mathbf{q}}$ is the angle between $\mathbf{q}$ and $\hat{x}$ axis, $\rho$ is the mass density of the ion, and the sound velocity $v_s^\mu$ is defined in terms of the only two independent elastic tensor coefficients $C_1$ and $C_2$. The quantized displacement field can then be expressed in terms of bosonic operators as
\begin{align}\label{eq:quantizedphonon}
	\tilde{\textbf{u}}_{\textbf{q}}^\mu(t) = i\sqrt{\frac{\hbar}{2\rho\delta_V\Omega_\textbf{q}}}\pa{a_\textbf{q}e^{-i\Omega_\textbf{q}t} + a^\dagger_{-\textbf{q}}e^{i\Omega_\textbf{q}t} },
\end{align}
where $\delta_V$ is the unit cell area.



\subsection{Spin-phonon interaction}\label{sec_spin-phham}
The magnetoelastic coupling arises from the variation in the Kitaev exchange strength $J$ due to lattice deformations. Assuming $J$ depends only on the distance $\textbf{r}$ between neighboring ions and $\textbf{u}(\textbf{r})$ is a small displacement, we can expand the spin interaction around the equilibrium value:
 $J_\alpha \rightarrow J_\alpha + \nabla J_\alpha \cdot \pa{\textbf{u}(\textbf{r}) - \textbf{u}(\textbf{r} + \textbf{M}_\alpha)}$, where $\textbf{M}_\alpha$
 are the nearest neighbor vectors (see Fig.~\ref{fig:fig1}(a))). From this approximation, the spin-phonon coupling Hamiltonian is given by \cite{ye2020phonon,metavitsiadis2020phonon}:
\begin{align}\label{Hspin-phonon}
\begin{split}
	\mathcal{H}_{\text{spin-ph}} = \lambda\sum_{\textbf{r},\alpha}\textbf{M}_\alpha\cdot\co{(\textbf{M}_\alpha\cdot\nabla)\,
 \textbf{u}(\textbf{r})}\sigma^\alpha_\textbf{r}\sigma^\alpha_{\textbf{r}+\textbf{M}_\alpha},
\end{split}
\end{align}
where $\lambda \sim (\nabla J_\alpha)_{\text{eq}} \ell_a$ is the spin-phonon coupling strength and $\ell_a$ is the lattice constant.  The spin-phonon coupling Hamiltonian (\refeq{Hspin-phonon}) can be rewritten in terms of symmetry channels. The two contributions are the $A_1$ and $E_2$ channels. Since the $E_2$ channel is dominant~\cite{ye2020phonon}, we will disregard the $A_1$ contributions here \cite{ye2020phonon,feng2021temperature}.  In the clean model, the contribution to $\mathcal{H}_{\text{spin-ph}}$ from the $E_2$ channel can be  explicitly written in terms of  Majorana fermions as
\begin{widetext}
	\begin{align}\label{eq:couplingham_MAJ}
		\hspace{-3mm}\mathcal{H}_{\text{spin-ph}}^{E_2} = -i\lambda\sum_{\textbf{r}}&\co{ \pa{\epsilon_{xx} - \epsilon_{yy}} \pa{\eta_{\textbf{r},\textbf{r}+\textbf{M}_x}c_{\br,A}c_{\br+\textbf{M}_x,B} + \eta_{\textbf{r},\textbf{r}+\textbf{M}_x}c_{\br,A}c_{\br+\textbf{M}_y,B} - 2\eta_{\textbf{r},\textbf{r}+\textbf{M}_x}c_{\br,A}c_{\br+\textbf{M}_z,B}}} +\\
		&  + 2\sqrt{3}\epsilon_{xy} \pa{\eta_{\textbf{r},\textbf{r}+\textbf{M}_x}c_{\br,A}c_{\br+\textbf{M}_x,B} - \eta_{\textbf{r},\textbf{r}+\textbf{M}_x}c_{\br,A}c_{\br+\textbf{M}_y,B} }. \nonumber
	\end{align}
\end{widetext}
This description extends easily to the case of quasivacancies. For any site in the quasivacancy space $\mathcal{Q}$, the spin-phonon coupling changes from $\lambda$ to $\lambda' \sim (\nabla J'_\alpha)_{\text{eq}} \ell_a$, where $J'_\alpha$ is the weak coupling between a quasivacancy and a normal spin. Thus, the magnetoelastic coupling Hamiltonian involving a magnetic defect retains the same functional form as
 Eq.~\eqref{eq:couplingham_MAJ}.

The next step is to derive interaction between phonons and Majorana fermions from the previously obtained magnetoelastic coupling. We use a mixed representation 
\cite{feng2021temperature},   where the displacement field is in the reciprocal space of the phonons, and the Majorana operators are in real space. By performing a Fourier transform on the strain tensor
 $\epsilon_{ij}(\textbf{r}) = \sum_\textbf{q}\frac{i}{2}\pa{q_iu_{\textbf{q},i} + q_ju_{\textbf{q},j}}$ and changing the basis, we express the coupling in terms of the displacement vectors
  $\tilde{\textbf{u}}_\textbf{q}^\mu$ in the longitudinal and transversal polarizations.   The spin-phonon coupling Hamiltonian in this mixed representation  can be written as 
  \begin{align}\label{eq:sph}
	\mathcal{H}_{\text{spin-ph}} = \frac{1}{\sqrt{N}}\sum_\textbf{q}V_\textbf{q}, 
\end{align}	
  where the interaction $V_\textbf{q}$ is explicitly given by
\begin{align}\label{eq:Vqmajs}
	V_\q = -\frac{i}{2} \sum_{\mean{i\in A,j\in B},\mu}c_ic_j \lamb{ij}\widetilde{u}_\q^\mu e^{i\q \cdot \textbf{r}_i}, 
\end{align}	
where $  \lamb{ij}$ define the matrix that couples the Majoranas at sites $\textbf{r}_{i,A}$ and $\textbf{r}_{j,B}$ with a phonon of momentum $\textbf{q}$.
In the  model with quasivacancies,  the elements of the  coupling vertex matrix $\lamb{}$ are given by 
\begin{align}
	\lamb{ij} = 
	\begin{cases}
		&2i\lambda \eta^\alpha_{ij} f^\mu_\alpha(\textbf{q}) \quad \text{if}\,  (i,j) \in \mathcal{P}\\
		&2i\lambda^\prime \eta^\alpha_{ij} f^\mu_\alpha(\textbf{q}) \quad \text{if either}\,  i 
		\, \text{or}\, j \in \mathcal{Q},
	\end{cases}
\end{align}
where $\mathcal{P}$ and $\mathcal{Q}$ are the normal site and quasivacancy subspaces, respectively. The explicit form of the functions  $f^\mu_\alpha(\q)$ is presented in Appendix \ref{App:MFP_vertex}.\\

It is important to note that the matrix  $\lamb{}$ is defined with entries from sublattice $A$ to sublattice $B$ ($i \in A$ and $j \in B$). Therefore, the matrix $\lamb{}$ must have the same structure as the adjacency matrix, where the entries are defined by the polarization, flux sector, phonon momentum $\textbf{q}$, and quasivacancy distribution. Similar to the Hamiltonian matrix, for a given quasivacancy, the procedure of finding the coupling vertex involves substituting $\lambda \rightarrow \lambda'$ for the corresponding row and column.

Since the calculation is performed on the whole lattice, it is useful to write Eq.~\eqref{eq:Vqmajs} in matrix form in the Majorana basis  $\boldsymbol{c} = (c_A \, c_B)$, i.e., in the sublattice representation. 
However, we cannot simply absorb the phase factor into the definition of $\lamb{AB}$ because $\lamb{BA} = \co{\lamb{AB}}^\dagger$,  which would change the phase factor from 
 $e^{i\textbf{q}\cdot\textbf{r}_j}$ to $e^{-i\textbf{q}\cdot\textbf{r}_j}$.  To incorporate all spatial dependence into a single compact notation, we introduce the matrix  $\Lamb{}$, defined by absorbing the exponential factor in \eqref{eq:Vqmajs} as the Hadamard product of $\lamb{ij}$ with the matrix $\mathbb{E}_\textbf{q}$, which is the symmetrized matrix with all the phase factors stored in rows \cite{feng2021temperature} (see Appendix \ref{App:MFP_vertex} for details):
\begin{align}
	\Lambda_{\q}^\mu = 	\begin{pmatrix}
		0 & \lambda_{\q}^\mu \\
		-\co{\lambda_{\q}^\mu}^T& 0 
	\end{pmatrix} \odot
	\begin{pmatrix}
		0 & \mathbb{E}_\q \\
		\co{\mathbb{E}_\q}^T& 0 
	\end{pmatrix}.
\end{align}
This allows  us to obtain  the coupling $V_\q$ in the sublattice representation:
\begin{align}\label{eq:Vq_matrix}
	V_\q = -\frac{i}{2}
	\begin{pmatrix}
		c_A & c_B
	\end{pmatrix}
	\begin{pmatrix}
		0 & \Lamb{AB}\\
		\Lamb{BA}& 0
	\end{pmatrix}
	\begin{pmatrix}
		c_A\\c_B
	\end{pmatrix}\widetilde{u}_\q^\mu,
\end{align}
with
\begin{align}
\begin{split}
	&\co{\Lambda_{\q,AB}^{\mu}}_{ij} = \lamb{ij}\; \frac{1}{2}\pa{e^{i\q\cdot\textbf{r}_i} + e^{i\q\cdot\textbf{r}_j}} \label{Lamb_matelem},\\
	&\co{\Lambda_{\q,BA}^{\mu}}_{ij} = -\lamb{ji} \;  \frac{1}{2}\pa{e^{i\q\cdot\textbf{r}_j} + e^{i\q\cdot\textbf{r}_i}}.
\end{split}
\end{align}
An immediate advantage of this expression is that
$\Lambda_{\q,BA}^{\mu,s}$ can be written in terms of $\Lambda_{\q,AB}^{\mu,s}$ as
\begin{align}\label{eq:lambda_BAtoAB}
	&\Lambda_{\q,BA}^{\mu,s} = -\co{\lambda_\q^{\mu} \odot \mathbb{E}_\q}^T  = -\co{\Lambda_{\q,AB}^{\mu,s}}^T.
\end{align} 
This form of the vertex matrix greatly simplifies the numerical computation of the polarization bubble,  as the vertex from $B$ to $A$ is determined once we know $\Lamb{AB}$.\\

%
%

\section{Phonon dynamics in the disordered system}\label{sec_phonon_dyn}


In this section, we present the computation of the phonon self-energy in the spin-phonon coupled Kitaev model with quasivacancies. In order to study the leading order corrections to phonon dynamics due to the spin-lattice coupling, we compute the one-loop phonon self-energy $\Pi^{\mu\nu}(\q,\Omega)$. As we will see in the next section, the sound wave attenuation coefficient $\alpha_s(\textbf{q})$ can be related to the imaginary part of the phonon polarization bubble.

Since we are interested in the phonon dynamics at finite temperatures, we compute the polarization bubble in real space using the Matsubara formalism. After deriving an expression for  $\Pi^{\mu\nu}(\textbf{q},\Omega)$ in terms of the Majorana fermion dispersion and the vertex matrix, we show the numerical results for the imaginary part of $\Pi^{\mu\nu}(\textbf{q},\Omega)$. Our analysis reveals how the localized modes induced by quasivacancies contribute to the self-energy in the low-energy regime.

\subsection{Phonon polarization bubble}\label{subsec:phonon_pol_bubble}
To calculate the phonon polarization bubble,  we need to rewrite the coupling vertex  (\refeq{eq:Vq_matrix})  in the basis of phonons and Bogoliubov quasiparticles. This involves relating the Bogoliubov quasiparticle operators (\refeq{bogtrans})  to the Majorana fermions through the following transformations:
\begin{align}
	\boldsymbol{c} = U\, W^\dagger \, \boldsymbol{\beta}, \quad \,\,
	\boldsymbol{c}^\dagger = \boldsymbol{\beta}^\dagger \, W\, U^\dagger, \label{ctobeta}
\end{align}
where $\boldsymbol{\beta}^\dagger = (\beta^\dagger \, \beta)$ and $\boldsymbol{c} = (\cb_A \, \cb_B)^T$. 
In this basis, the coupling $V_\q$ is written as:
\begin{align}
	V_\q = -\frac{i}{2}	\begin{pmatrix}
		\beta^\dagger & \beta
	\end{pmatrix}
	\begin{pmatrix}
		\widetilde{\Lambda}^\mu_{\q,11} & \widetilde{\Lambda}^\mu_{\q,12}\\
		\widetilde{\Lambda}^\mu_{\q, 21} & \widetilde{\Lambda}^\mu_{\q,22}\\ 
	\end{pmatrix}
	\begin{pmatrix}
		\beta \\ \beta^\dagger
	\end{pmatrix}\widetilde{u}_\q^\mu,  \label{Vqbogo}
\end{align}
where the vertex
\begin{align}
	\widetilde{\Lambda}_{\q}^\mu = WU^\dagger\Lambda_\q^\mu U W^\dagger \label{lambtildef}
\end{align}
is written in a block form in (\refeq{Vqbogo}), corresponding to 
the creation and annihilation subspaces. In this formulation,  $\widetilde{\Lambda}^\mu_{\q,11}$ and $\widetilde{\Lambda}^\mu_{\q,22}$  contribute to the particle-hole (ph-) channel, while the off-diagonal components contribute to the particle-particle (pp-) channel. 
 

In the Matsubara formalism, the phonon polarization bubble can be written as the following expectation value:
\begin{align}\label{Pi_def}
	\Pi^{\mu\nu}(\q,\tau) = \expval{T_\tau\left\{\pa{ \betab^\dagger \tL{\mu}{\q}\betab }(\tau)
		\pa{ \betab^\dagger \tL{\nu}{-\q}\betab }(0)\right\}},
\end{align}
where  $T_\tau$ is the imaginary time-ordering operator with $\tau = it$. As shown in Appendix \ref{App:steps_PI}, the  Fourier transform of $\Pi^{\mu\nu}(\q,\tau)$ is  given by 
\begin{align}\label{eq:piomega_full}
	\Pi^{\mu\nu}(\q,\Omega) = 
	&\frac{1}{N}\sum_{ij}
	P^{\bar{g}g}_{ij}\co{\tLambij{\mu}{11}\odot\pa{\tLambmij{\nu,T}{11} - \tLambmij{\nu}{22}} }_{ij} \nonumber\\
	&+P^{g\bar{g}}_{ij}\co{\tLambij{\mu}{22}\odot\pa{ \tLambmij{\nu,T}{22} - \tLambmij{\nu}{11} }}_{ij} \nonumber\\
	&+P^{gg}_{ij}\co{\tLambij{\mu}{21}\odot\pa{\tLambmij{\nu,T}{12} - \tLambmij{\nu}{12} }}_{ij} \nonumber\\
	&+P^{\bar{g}\bar{g}}_{ij}\co{\tLambij{\mu}{12}\odot\pa{[\tLambmij{\nu,T}{21} -\tLambmij{\nu}{21}}}_{ij},
\end{align}
where the sum is over all the eigenstates $\varepsilon_i$ and $\varepsilon_j$ and $P^{\bar{g}g}_{ij}$, $P^{g\bar{g}}_{ij}$, $P^{gg}_{ij}$ and $P^{\bar{g}\bar{g}}_{ij}$ are the convolutions of the Matsubara Green's functions, which are explicitly written as:
\begin{align}
\begin{split}
		&P_{ij}^{g\bar{g}} = \frac{n_F(\varepsilon_i) - n_F(\varepsilon_j)}{\Omega - \varepsilon_i + \varepsilon_j +i\delta}\label{Pggbar} \\ 
		&P_{ij}^{\bar{g}g} = \frac{n_F(-\varepsilon_i) - n_F(-\varepsilon_j)}{\Omega + \varepsilon_i - \varepsilon_j+i\delta}  \\ 
		&P_{ij}^{gg} = \frac{n_F(\varepsilon_i) - n_F(-\varepsilon_j)}{\Omega - \varepsilon_i - \varepsilon_j+i\delta}  \\
		&P_{ij}^{\bar{g}\bar{g}} = \frac{n_F(-\varepsilon_i) - n_F(\varepsilon_j)}{\Omega + \varepsilon_i + \varepsilon_j+i\delta}
\end{split}
\end{align}
Here, $P^{gg}_{ij}$ contributes to the pp-channel, while $P^{\bar{g}g}_{ij}$ and $P^{g\bar{g}}_{ij}$ contribute to the ph-channel. The terms with $P^{\bar{g}\bar{g}}_{ij}$ are in the hole-hole (hh-) channel, which will not be considered in our calculations. Also, we point out that the temperature dependence of the polarization bubble is encoded in the expressions for each $P_{ij}$ through the Fermi-Dirac distribution. On the other hand, the vertex matrices $\tLamb{ij}$ determine the angular dependence of the Majorana fermion-phonon scattering.\\


\subsection{Numerical results for the phonon self-energy}\label{sec:Numerical_self-energy}

\begin{figure*}
	\centering
	\includegraphics[width=1\linewidth]{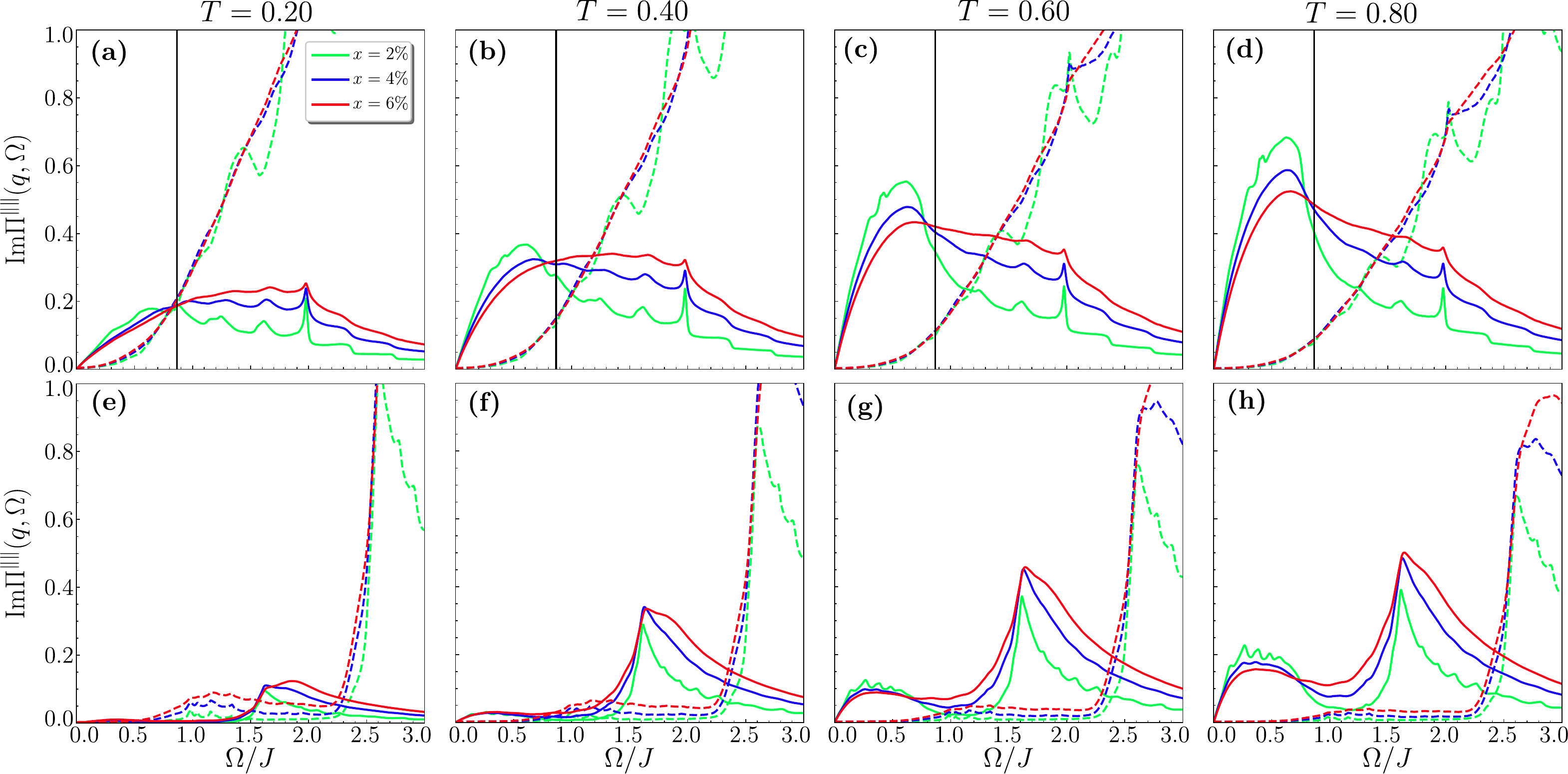}
	\caption{ Imaginary part of the phonon self-energy calculated in the zero-flux sector. The pp-channel (dashed lines) and ph-channel (solid lines) are plotted as a function of the phonon frequency $\Omega$ with $|\textbf{q}| = 0.5$ and $\theta_{\mathbf{q}} = \pi/6$ and different values of temperature. The first row ((a)-(d)) corresponds to $\kappa = 0.0$ and the second ((e)-(h)) to $\kappa = 0.2$. Each curve is averaged over 10000 disorder realizations with system size $L = 32$ with $J^\prime = 0.01J$, under a Lorentzian broadening $\delta = 0.1$. The three colors correspond to different values of quasivacancy concentrations (see the label in (a)). The vertical line in the first row indicates the energy $\Omega = v_F q$. Both $\Omega$ and $T$ are in units of $J$.}
	\label{fig:implot0flux}
\end{figure*}


In this section, we present our results for the phonon self-energy as a function of the input phonon frequency  $\Omega$. We restrict our analysis to the long-wavelength approximation, calculating $\Pi^{\mu\mu}_{\text{ph}}(\textbf{q},\Omega)$ for small values of $\textbf{q}$  and,  correspondingly, $\Omega$. 

The imaginary part of the self-energy $\Im [\Pi^{\parallel\parallel}_{\text{ph}}(\textbf{q},\Omega)]$, shown in 
Fig.~\ref{fig:implot0flux} as a function of phonon frequency  $\Omega$ at different temperatures $T$ and concentration of quasivacancies $x$, describes the spectral width of the phonon, providing insights into its lifetime and sound attenuation.
In the first row of Fig.~\ref{fig:implot0flux},  we plot $\Im [\Pi^{\parallel\parallel}_{\text{ph}}(\textbf{q},\Omega)]$  for the time-reversal invariant case, $\kappa=0$,
and in the second row for $\kappa=0.2$.

In the clean limit, due to strong kinematic constraints, the pp-channel is restricted to $v_s>v_F$,  while the ph-channel is only allowed when $v_s<v_F$. Since we fix the value of 
 $|{\bf q}|=0.5$, the vertical line in the first row of Fig.~\ref{fig:implot0flux} separates these regions.
There, 
we observe that the ph-channel dominates at small $\Omega$, with its contribution increasing with temperature due to higher occupation of low-energy states and thus increasing phase space. This also leads to a distinctive linear  $T$ dependence of the sound attenuation for $v_s<v_F$. 
Conversely, the pp-channel becomes more significant at higher $\Omega$.

In the presence of defects, the translation symmetry is lost so the Majorana fermion momentum $\textbf{k}$ is not a good quantum number anymore. Therefore, at a finite concentration of quasivacancies, the momentum constraint is lifted and the decay of a phonon involves both the ph- and pp-processes.
Additionally, the Fermi velocity $v_F$ is not well defined anymore, due to the disappearance of the Dirac cones. However, we can still take $v_F$ as a parameter to define an energy scale vestigial to the clean limit. In the next section, we will see how the attenuation changes for $v_s/v_F$ larger or smaller than 1.

 In the finite-field case (second row of Fig.~\ref{fig:implot0flux}), we observe the formation of a peak structure at low-$\Omega$ in both channels. This is a direct consequence of the low-energy modes induced by vacancies, as discussed in Sec.~\ref{sec_spinham}. At very low temperatures, there is no contribution from the 
 ph-channel because the low-energy states are gapped out. However, as the temperature increases, the in-gap Majorana modes start to become occupied. This occupation modifies the manifold of allowed scattering processes, resulting in the previously mentioned peak structure. In the following section, we will explore how this influences the evolution of sound attenuation as a function of temperature.\\




\section{Sound attenuation}\label{sec_soundatt}



The attenuation process of an acoustic sound wave within the material can be described quantitatively by considering a lossy acoustic wave function that decays with distance from the driving source
\begin{align}
	\textbf{u}(\textbf{r},t) = \textbf{u}_0 e^{-\alpha_s(\textbf{q})r}e^{i(\Omega t - \textbf{q}\cdot \textbf{r})},
\end{align}
where $\textbf{u}$ is the displacement field with $\textbf{u}_0 = \textbf{u}(t=0)$, $\Omega$ is the acoustic wave dispersion and $\textbf{q}$ is the wave-vector. The sound attenuation coefficient,  $\alpha_s(\textbf{q})$, is related to the diagonal component of the imaginary part of the phonon self-energy \cite{ye2020phonon}:
\begin{align}
	\alpha^\mu_s(\textbf{q}) \propto -\frac{1}{(v^\mu_s)^2 q}\Im[\Pi^{\mu\mu}(\textbf{q},\Omega)]_{\Omega = v_sq}.
\end{align}

 As we already discussed above, the phonon dynamics in the pure model is strongly dependent on the relative values of the sound velocity $v_s$ and the Fermi velocity $v_F$. In the case when $v_s<v_F$, earlier studies have demonstrated that $\alpha_s(\textbf{q})$ exhibits a particular  sixfold angular dependence on $\textbf{q}$ and shows a linear dependence on temperature at low-$T$ \cite{ye2020phonon}:
\begin{align}
	\alpha^{\parallel}_s(\textbf{q}) \sim \frac{\lambda^2 v_sq}{v_F^3 } T \, (1-\cos 6\theta_\textbf{q}).
\end{align}
The behavior of the transversal polarization can be obtained from a simple $\pi/2$ rotation from the expression above. Therefore, for simplicity, we only refer to the longitudinal component in the remainder.

The most crucial feature in this expression of $\alpha_s$, however, is the linear scaling with temperature. 
As argued in \cite{ye2020phonon,Lee2011PhysRevLett.106.056402}, this behavior is an indirect signature of the fractionalized nature of the spin excitations in this model, as it comes from to the phase space of the Majorana fermion-phonon scattering. 
Clearly, the anharmonic effects  due to the phonon-phonon scattering processes always contribute in a realistic material
 but their  contribution is
 $\sim T^3$. Consequently, at sufficiently low temperatures, the sound attenuation in the quantum spin liquid predominantly arises from the decay into fractionalized excitations.
 In $\alpha$-RuCl$_3$, there is also a contribution to the sound attenuation from scattering of the phonons from magnons. However, given that magnons are gapped at low temperatures, when  $\alpha$-RuCl$_3$ is magnetically ordered, and short-ranged above the Neel ordering temperature, their contribution on sound attenuation is negligible \cite{Hauspurg2024}. \\

\subsection{Temperature dependence of $\alpha_s(\textbf{q})$: Qualitative analysis}\label{sec_powercount}

Here we explore how the linear scaling with temperature can be derived using a straightforward power counting argument directly from the phonon polarization bubble. 
 While this derivation is more easily performed in a translationally invariant system within $\textbf{k}$-space,
  It can also be done
 within a real-space formulation.  
 By performing this analysis, we can more transparently connect the clean limit to our numerical calculations of sound attenuation in the disordered system, since whether in $\textbf{k}$-space or real-space, it highlights the role of the density of states $\rho(\varepsilon)$.

For illustration, we first consider the clean system in the zero-flux sector. Starting with the ph-channel, symmetry allows us to consider only the $g\bar{g}$ process, which contributes to sound attenuation as follows:
\begin{widetext}
	
	\begin{align}    
 \begin{split}
		\text{Im}\,\Pi^{\mu\mu}_{g\bar{g}}(\textbf{q},\Omega) &= \frac{1}{N}\text{Im}\sum_{ij}\frac{n_F(\varepsilon_i) - n_F(\varepsilon_j)}{(\Omega+i\delta) - \varepsilon_i + \varepsilon_j}
		\co{\tLambij{\mu}{11}\odot\pa{\tLambmij{\mu,T}{11} - \tLambmij{\mu}{22}} }_{ij}\\
		&\equiv -\frac{\pi}{N}\sum_{ij}\co{n_F(\varepsilon_i) - n_F(\varepsilon_j)}\, \delta\pa{\Omega - \varepsilon_i + \varepsilon_j} [\xi_{g\bar{g}}^\mu(\textbf{q})]_{ij}.
\end{split}	
 \end{align}
	\end{widetext}
Here, we used the explicit form of
 $P^{\bar{g}g}_{ij}$ from Eq.~\eqref{Pggbar},
 with
 $n_F(\varepsilon)$ being the Fermi-Dirac distribution, and we assume that $\nu=\mu$ from now on. Using the relation  $\Im\frac{1}{\Omega + i\delta} = -\pi\delta(\Omega)$, we  obtain the second line . We also define the product of vertex functions   ${\widetilde{\Lambda}}_\textbf{q}^\mu$ in the $g\bar{g}$ channel as $\xi_{g\bar{g}}^\mu(\textbf{q})$, for simplicity. It is crucial to note that the matrix elements of  $\xi^\mu_{g\bar{g}}(\q)$ weight the sum of Fermi distributions for different energy pairs $\varepsilon_i, \varepsilon_j$, which are related to $\Omega$ via the kinematic constraints. However, the idea of the concept of power counting is that the contribution from  $\xi^\mu_{g\bar{g}}(\q)$ does not significantly affect the temperature dependence of sound attenuation. Therefore, we can extract the scaling of $\alpha_s$ in $T$ solely by examining the remaining sum of Fermi distributions. 
 To make this approximation, we assume 
  $[\xi_{g\bar{g}}^\mu(\textbf{q})]_{ij} = \xi^{\mu}_{g\bar{g}}(\textbf{q})$, implying that the vertex function is constant with respect to the eigenvalue indices. This approach is equivalent to neglecting the 
  $\textbf{k}$ dependence of $\lamb{\textbf{k}}$ in the clean case \cite{ye2020phonon}.

As a result, we write the imaginary part of the self-energy as
\begin{align}
	\text{Im}\,\Pi^{\mu\mu}_{g\bar{g}}(\textbf{q},\Omega) &= \xi^\mu_{g\bar{g}}(\q)\widetilde{\Pi}_{g\bar{g}}(\Omega,T),
\end{align}
where the function $\widetilde{\Pi}(\Omega,T)$  encodes all the temperature dependence and is expressed in terms of the density of states $\rho(\varepsilon)$ as
\begin{align}
\begin{split}
&\widetilde{\Pi}_{g\bar{g}}(\Omega,T) =\frac{1}{N}\sum_{ij} \co{n_F(\varepsilon_i) - n_F(\varepsilon_j)}\delta\pa{\Omega - \varepsilon_i + \varepsilon_j} \\
&  = -\pi \int_0^{\varepsilon_{\mathrm{max}}} d\varepsilon \rho(\varepsilon)\rho(\varepsilon + \Omega)\, [n_F(\varepsilon + \Omega) - n_F(\varepsilon)],\label{PH_powercount}
\end{split}
\end{align}
where the upper bound $\varepsilon_{\mathrm{max}}$ is taken as the bandwidth $6J$. This estimation is justified in the clean limit when the temperature is low enough that only states around the Dirac points contribute,
$T\ll J$, and the vertex function depends only on
$|\textbf{q}|$, i.e.,  $\tL{\mu}{\textbf{q}} \sim |\textbf{q}|$. Therefore, we can approximate the dependence in $T$ by expanding the Fermi-Dirac functions in the limit where $\varepsilon,\varepsilon + \Omega \ll T$. This, together with the linear density of states  of  the Dirac cone yields
\begin{align}
\begin{split}
\widetilde{\Pi}_{g\bar{g}}(\Omega,T) &= -\pi\int_0^{T} d\varepsilon\, (\varepsilon^2 + \Omega\varepsilon)\pa{\frac{1}{2}-\frac{\varepsilon+\Omega}{4T} - \frac{1}{2} + \frac{\varepsilon}{\Omega}}\\
& = \frac{\pi}{8}\Omega^2 T + \mathcal{O}(T^2).
\end{split}
\end{align}

It can be easily shown that the $\bar{g}g$ process contributes in the exact same way. Therefore, our real-space calculation reproduced the result reported in \cite{ye2020phonon}.

The power-counting estimation in pp-channel can be performed by following the same procedure for the $gg$-process. Then, 
\begin{align}
\widetilde{\Pi}_{gg}(\Omega,T) &=  \pi\int_0^\Omega d\varepsilon \, \rho(\varepsilon)\, \rho(\Omega - \varepsilon)\co{n_F(\varepsilon) - n_F(\varepsilon-\Omega)},
\end{align}
 where the upper bound in this case is $\Omega$ due to the restriction set by $\rho(\Omega-\varepsilon)$.
Because the pp-processes can happen at any temperature, by just taking $T\rightarrow 0$ limit of the Fermi distributions,
 we estimate $\widetilde{\Pi}_{gg}(\Omega,T)$  as
\begin{align}
	\widetilde{\Pi}_{gg}(\Omega,T\rightarrow 0) \approx \pi\int_0^\Omega d\varepsilon \, \rho(\varepsilon)\, \rho(\Omega - \varepsilon)
	\label{PP_powercountT0}
\end{align}
Therefore, the sound attenuation at very low temperatures is strongly dependent on the form of the density of states. On the other hand, we can expand the integrand at large temperatures to obtain the characteristic $1/T$ dependence of the pp-channel when $T\gg \Omega$: 
\begin{align}
	\widetilde{\Pi}_{gg}(\Omega,T) \approx \frac{\Omega}{4T}\, \widetilde{\Pi}_{gg}(\Omega,0).\label{PP_powercountT}
\end{align}
Thus, the pp-channel decays as $1/T$  even in the disordered systems where the low-energy density of states deviates from the linear form.

To summarize, our power counting argument suggests that sound attenuation scales linearly with temperature at low $T$ if the ph-channel dominates, and as 
 $\sim 1/T$
 at high $T$ if the pp-channel is prevalent. 
In the site-disordered system, the low-energy density of states is more intricate so the qualitative argument can hardly be applied.
 Therefore, in the next section, we will use numerical results to accurately scale $\alpha_s(T)$ in the Kitaev spin liquid with quasivacancies. We will show that even in this case, $\alpha_s(T)$ only moderately deviates from the estimates in the clean case due to the distinct fermionic nature of the scattering processes in the quantum spin liquid.



\subsection{Quasivacancies: numerical results }\label{sec_soundatt_Quasivac}

\begin{figure*}[t]
	\centering
	\includegraphics[width=1\linewidth]{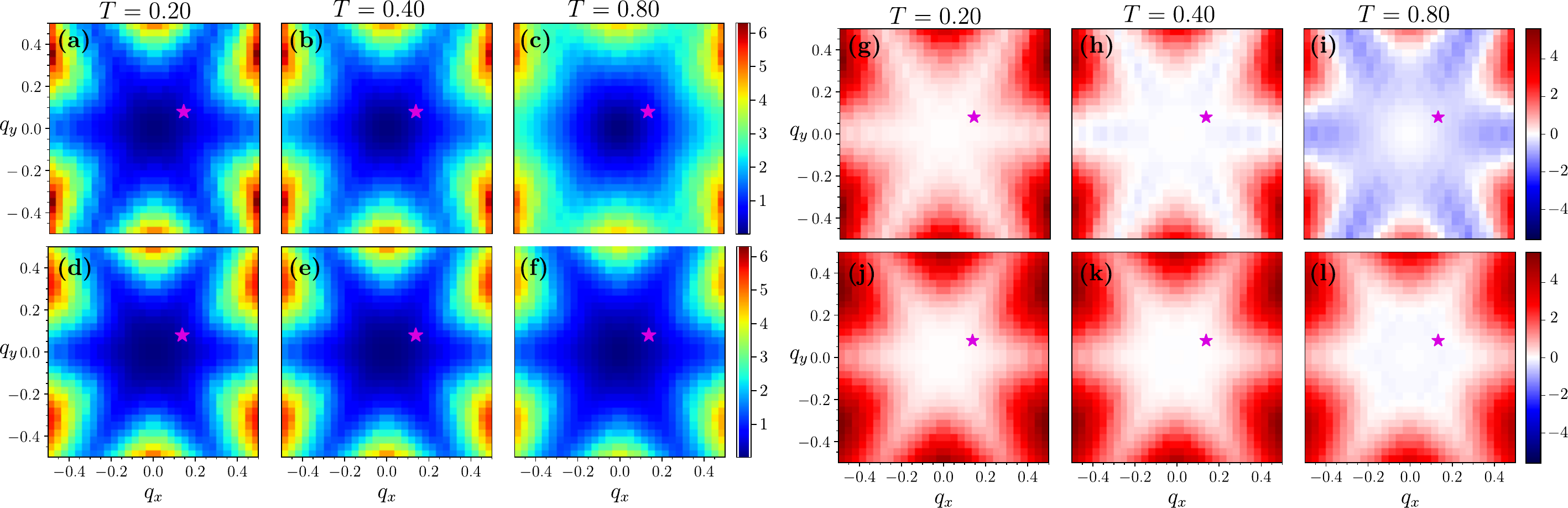}
	\caption{
 (a)-(f) Sound attenuation $\alpha_s^\parallel(\textbf{q})$ for $x = 4\%$ in the zero-Flux sector at $T/J = 0.2, 0.4, 0.8$, with phonon momenta in the region  $q_x,q_y \in [-\pi/2,\pi/2]$. (g)-(l) Difference between the pp- and ph-contributions to the sound attenuation, with a positive sign indicating dominant pp-contribution and the negative sign indicating dominant ph-contribution. The first row ((a)-(c) and (g)-(i)) shows results for $v_s/v_F = 0.8$, and the second row ((d)-(f) and (j)-(l)) for $v_s/v_F = 1.2$. We used $L = 24$ with $q_x \times q_y = 30\times 30 = 900$ points, and averaged over 100 disorder realizations.  We set  $J^\prime = 0.01J$ and broadening $\delta = 0.1$. The magenta star marks $|\textbf{q}| = 0.5$, $\theta_\textbf{q} = \pi/6$, used in the temperature evolution computation (see Fig.~\ref{fig:fig4}).
 }
	\label{fig:gridplot1}
\end{figure*}

In this section, we present our results for the sound attenuation coefficient in the presence of quasivacancies. We relax the kinematic constraints due to the lack of translation symmetry, imposing only energy conservation in each scattering channel. Consequently, we include both pp- and ph-contributions regardless of the relative sound velocity.

 We start by evaluating sound attenuation coefficient $\alpha^{\parallel}_s(\textbf{q})$ in the zero-flux sector. The left panel in Fig.~\ref{fig:gridplot1}  shows the angular dependence of $\alpha^{\parallel}_s(\textbf{q})$ for
  $\kappa = 0.0$ computed at $T/J = 0.2, \, 0.4, \, 0.8$ for both $v_s>v_F$ and $v_s<v_F$. Notably, as temperature increases, a distinction between 
 $v_s>v_F$ and $v_s<v_F$ emerge, even when summing over both scattering channels. This is further illustrated in the right panel of Fig.~\ref{fig:gridplot1}, which shows the difference between pp- and ph-contributions for a region of $\mathbf{q}$ close to the center of the Brillouin zone. For $v_s/v_F=0.8$,  due to dominant ph-processes, 
 the magnitude of $\alpha_s^\parallel(\textbf{q})$ increases as the fermionic population grows with temperature. For $v_s/v_F=1.2$, both channels contribute, with pp-scattering dominating, resulting in sound attenuation that is almost temperature-independent for small momentum $q$.
 We also observe that the sixfold symmetry remains intact, as expected due to the $C_6$ symmetry in the phonon sector and the averaged $C_3$ symmetry in the spin sector. This averaging is a result of preserving an equal number of quasivacancies on the A and B sublattices, despite their random positions.

%

Now we turn our attention to the temperature evolution of  $\alpha_s^\parallel(\textbf{q})$. Again,
we focus on the zero-flux sector, as the results for the bound-flux sector are very similar  (see Appendix \ref{App:bound-flux}). 
 Fig.~\ref{fig:fig4} shows the sound attenuation coefficient for $\kappa = 0.0$, where the characteristic scaling $\alpha_s^\parallel(\textbf{q}) \sim T$ is evident even in the presence of quasivacancies. This linear scaling with temperature occurs only at low $T$, as shown in the inset of Fig.~\ref{fig:fig4},  and this temperature range aligns with the linear in $T$ regime observed for some acoustic modes in ultrasound experiments in \rucl ~\cite{Hauspurg2024}.

We also point out that the sharp difference between the clean limit and all $x \neq 0$ curves at very low $T$ is caused by an almost zero-energy peak induced by quasivacancies, which allows more states to participate in the attenuation process. This is evident from the form of the pp-contribution in Eq.~\eqref{PP_powercountT0}, which is strongly dependent on the density of states.
At higher temperatures, we observe that $\alpha_s^\parallel(\textbf{q})$ decays, a trivial consequence of the Pauli exclusion principle, as most of the fermionic states become occupied. 
\\

\begin{figure}
	\centering
	\includegraphics[width=0.9\linewidth]{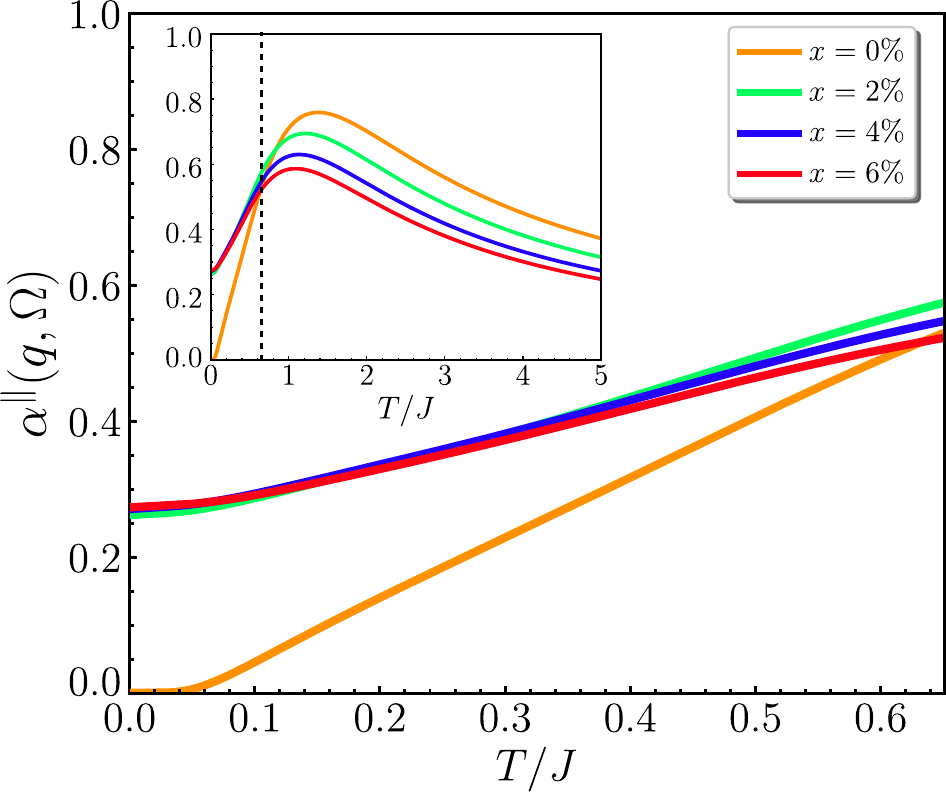}
	\caption{Sound attenuation as a function of temperature computed for $v_s/v_F = 0.8$. Inset:  The same plot for a broader range of temperatures. The calculations are done on the lattice with $L = 32$ and averaged over 1000 disorder realizations, along with the weak coupling $J^\prime = 0.01J$ and a broadening $\delta = 0.1$. The dashed line indicates the upper bound shown in the main plot. For the clean limit ($x=0\%$) we subtracted the pp-contribution. }
	\label{fig:fig4}
\end{figure}


%

\begin{figure}
	\centering
	\includegraphics[width=1\linewidth]{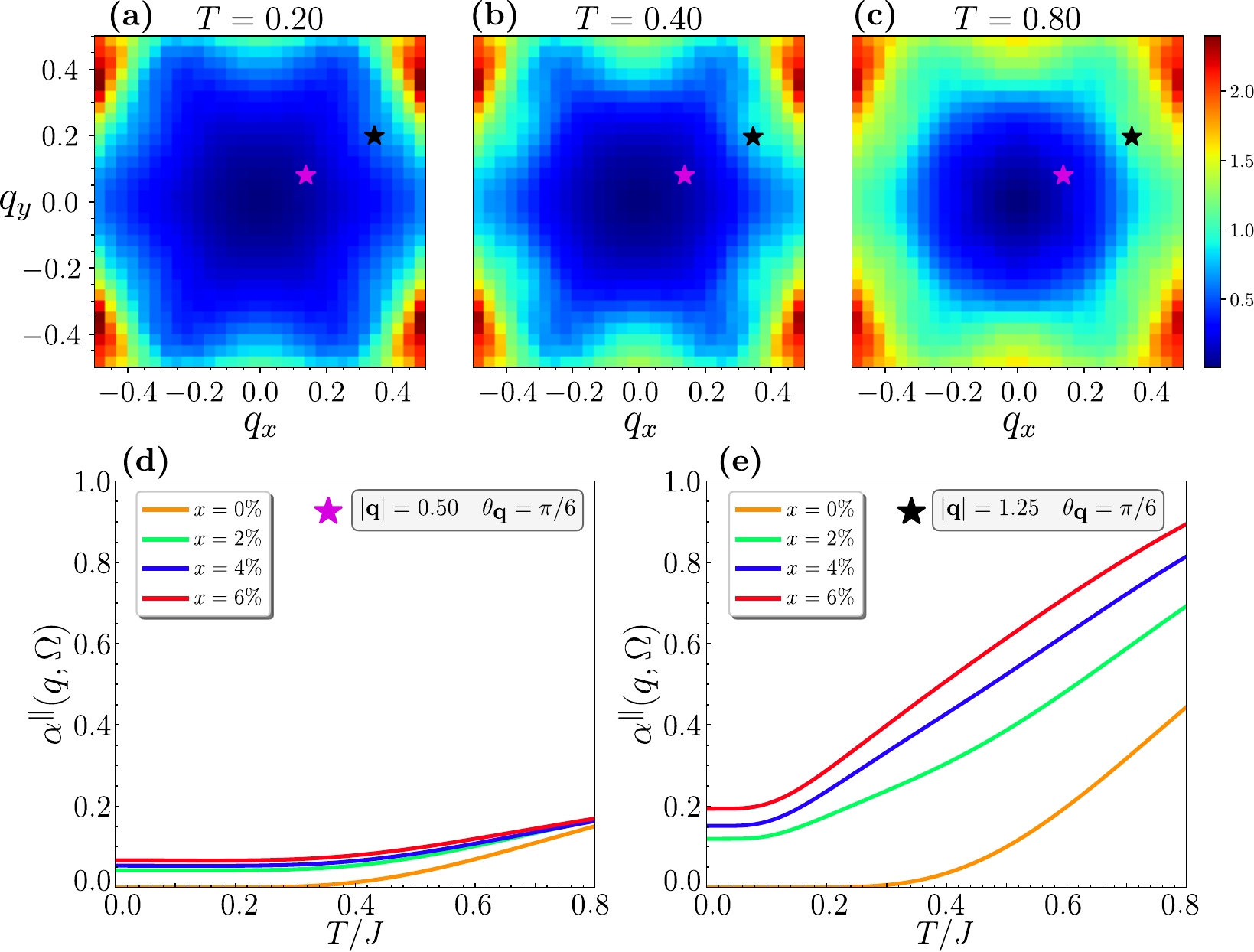}
	\caption{(a)-(c)Sound attenuation $\alpha_s^\parallel(\textbf{q})$ for $T = 0.4 J$, $v_s/v_F = 0.8$ and $\kappa = 0.2$ in the zero-Flux sector with $x =  4\%$ of quasivacancies. The phonon momenta are defined in the region  $q_x,q_y \in [-\pi/2,\pi/2]$. We use $L = 24$ with $q_x \times q_y = 30\times 30 = 900$ points averaged over 100 disorder realizations and $J^\prime = 0.01J$. The spectral broadening $\delta = 0.1$ is applied. In the second row we show $\alpha^{\parallel}(\textbf{q})$ as a function of temperature for different values of $x$ and $|\textbf{q}| = 0.5$ (d) and $|\textbf{q}| = 1.25$ (e) (both with $\theta_\textbf{q} = \pi/6$). These points are marked in the first row of the figure as magenta and black stars, respectively.}
	\label{fig:fig5}
\end{figure}

Finally, we present our results for the case $\kappa \neq 0$ in Fig.~\ref{fig:fig5}. The low-energy modes induced by quasivacancies inside the gap caused by $\kappa\neq 0$  significantly affect the sound attenuation as a function of $|\textbf{q}|$, as shown in Fig.~\ref{fig:fig5}(a)-(c) for different values of temperature and a fixed concentration $x=4\%$. 
Here we plot only the contribution from the ph-channel scattering, as it dominates at low $\Omega$ (see the second row of Fig.~\ref{fig:implot0flux}). At the center of the Brillouin zone, there is a region where $\alpha_s \ll 1$ due to the gap induced by $\kappa$. However, for large enough $|\textbf{q}|$, the ph-processes become allowed, indicating the scattering from in-gap states to the bulk. As temperature increases, the attenuation from these processes becomes more pronounced. This is evident  in Fig.~\ref{fig:fig5}(d)-(e), 
where we compare the temperature scaling for two distinct values of $\textbf{q}$, keeping the sound velocity 
 $v_s$ and  angle $\theta_{\mathbf{q}}$  fixed. Therefore, the phonon energy
  $\Omega_{\mathbf{q}}$ differs only by the choice of $|\textbf{q}|$. These points are depicted as the black and magenta stars in Fig.~\ref{fig:fig5}(a)-(c).
  For small  $|\textbf{q}|$ (Fig.~\ref{fig:fig5}(d)), we observe an $x-$independent upturn at moderate temperatures, indicating that only states with energies larger than
  the gap participate in the ph-processes.
In contrast, for a larger value of $|\textbf{q}|$ (Fig.~\ref{fig:fig5}(e)), 
we see an $x$-dependent upturn at much smaller temperatures, 
highlighting the contribution from localized states induced by the quasivacancies to the sound attenuation.
Interestingly, the linear temperature scaling persists even as the manifold of low-energy states changes dramatically, from a Dirac cone in the clean model to an in-gap peak at finite $\kappa$ with quasivacancies. These results are consistent with our power-counting argument.

\subsection{Flux disorder: Numerical results}\label{sec_soundatt_RF}

So far, we have focused on the scenario where the phase space for the scattering of acoustic phonons is entirely determined by the thermally excited Majorana fermions. This is justified by the form of the magneto-elastic coupling vertex (\ref{eq:Vq_matrix}), which in the Kitaev model is diagonal in the flux sectors. Nevertheless, thermally excited fluxes modify the spectrum of Majorana fermions and, therefore, indirectly affect sound attenuation. We will study this effect in this section.

As fluxes proliferate at finite temperatures, the bond variables $\eta_{ij}$ exhibit a temperature-dependent flipping probability distribution~\cite{feng2021temperature,kao2021localization,halasz2019observing}.
This distribution can be accounted for either by numerically costly Monte Carlo sampling of flux excitations~\cite{Nasu2014,feng2021temperature}, or alternatively, by a quantitative approximation of the finite-temperature behavior obtained by taking a random average over “typical” flux sectors~\cite{kao2021localization,halasz2019observing}.
In the random-flux regime, however, each plaquette has an equal probability of hosting either a zero-flux or a $\pi$-flux.
As shown in Fig.~\ref{fig:fig1}(e,f), this type of disorder flattens the Majorana fermion density of states across the entire energy range, allowing us to employ the power-counting argument in a straightforward manner.

By assuming a constant density of states, $\rho(\varepsilon) \sim \rho$, we can easily perform the integration in Eq.~\eqref{PH_powercount} to obtain the ph-channel contribution to the temperature dependence. Taking $\varepsilon_{\text{max}} \sim J$, we can write:
\begin{align}
	\widetilde{\Pi}_{g\bar{g}}(\Omega,T)=\pi \rho^2T\log\pa{1+\tanh\pa{\frac{J}{2T}}\tanh\pa{\frac{\Omega}{2T}}},
\end{align}
At high temperatures  ($T \gg J, \Omega$), we can recover the expected  $1/T$ behavior from the Pauli exclusion: 
$
\widetilde{\Pi}_{g\bar{g}} \sim \pi\rho^2 \frac{J\Omega}{4T}
$.
For small $T$ ($T \ll \Omega$),
we can expand the term inside the logarithm, allowing us to recover the linear in $T$  behavior:
$\widetilde{\Pi}_{g\bar{g}}\sim \pi \rho^2T\log 2$. By adding this to the low-temperature constant contribution from the pp-channel, we get the general behavior: $\widetilde{\Pi}(\Omega,T) \sim \alpha_0 + \alpha_1T$, where the constants $\alpha_0$ and $\alpha_1$ depend on the density of states and the frequency $\Omega$.

\begin{figure}
	\centering
	\includegraphics[width=0.9\linewidth]{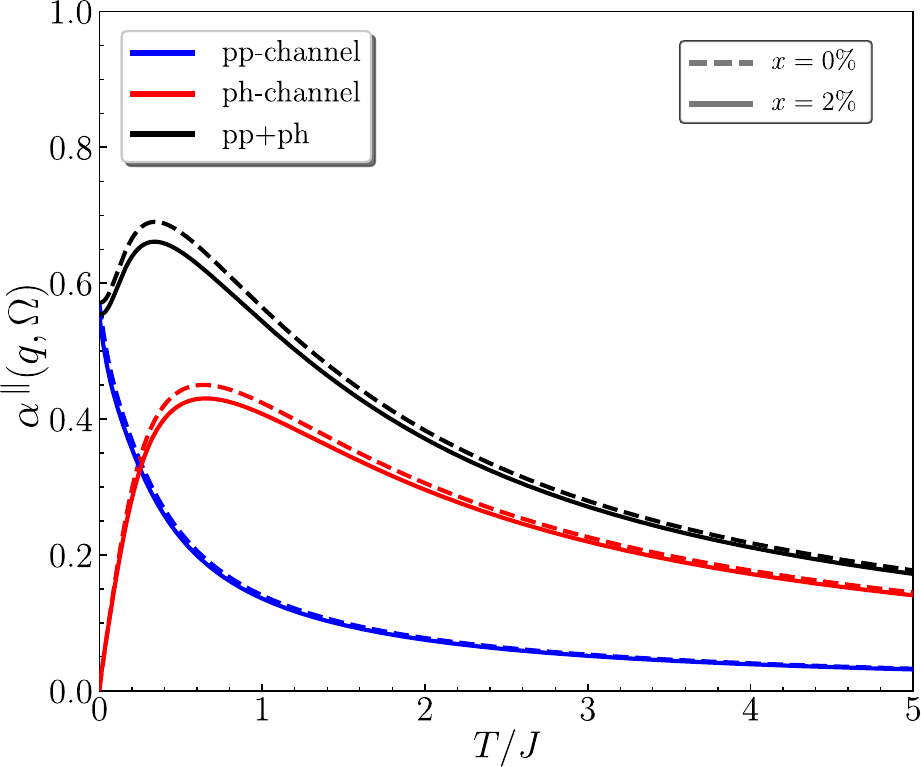}
	\caption{ The contributions of pp- and ph-channels, as well as their combined effect,  to the sound attenuation as a function of temperature in the random-flux sector. Inset: The same plot in the low-temperature region. In both figures we have $\kappa = 0$, $v_s/v_F = 0.8$, $L = 32$ with 1000 disorder realizations, $J^\prime = 0.01J$, and a broadening $\delta = 0.1$. The dashed line corresponds to the random-flux sector without quasivacancies, while the solid line indicates the case with both random flux and quasivacancies.}
	\label{fig:rf_num}
\end{figure}

This simple argument is corroborated by our numerical calculations, as shown in  Fig.~\ref{fig:rf_num}. Here we show the contributions of pp- and ph-channels, as well as their combined effect, to the sound attenuation for both the clean case and the case with $2\%$ of quasivacancies,
computed for  $\kappa = 0$. To this end, we plot the sound attenuation as a function of temperature in the random-flux sector. 
We clearly see that the overall behavior of $\alpha_s(\textbf{q})$ is almost independent of the presence of quasivacancies, as flux disorder overrides the contribution from the vacancy-induced low-energy localized modes.


%
%
%
%

%
%
\section{Conclusions}\label{conclusions}

In this work, we have investigated the effects of site disorder on the phonon dynamics of the Kitaev spin liquid on a honeycomb lattice.  Our primary objective was to determine the sound attenuation coefficient as a function of temperature in the presence of site disorder. We demonstrated that a linear temperature scaling may persist even with disorder. To achieve this, we extended the methods developed in previous studies on the acoustic phonon dynamics in the Kitaev model \cite{ye2020phonon,feng2021temperature} by incorporating quasivacancies into the model.
Because these types of defects affect only the magnetic interactions, the lattice symmetries are preserved, allowing us to employ 
the mixed representation method~\cite{feng2021temperature} to couple a phonon with momentum $\textbf{q}$ to the Majorana fermions labeled by real-space indices $i$ and $j$, with energies obtained via exact diagonalization.
 By constructing the coupling vertex within this framework, we were able to compute the phonon self-energy in the disordered system. Our numerics showed how the low-energy modes induced by quasivacancies change the imaginary part of the phonon polarization bubble. The effects of such states are particularly evident in the time-reversal symmetry-breaking case, $\kappa \neq 0$, where the scattering from localized modes to the bulk led to a low-$\Omega$ peak inside the gap for the ph-channel. The flux dependence of these localized modes \cite{kao2021localization,kao2024dynamics} resulted in distinct peak structures for the bound-flux and zero-flux scenarios.  


The sound attenuation coefficient $\alpha^\mu_s(\textbf{q})$ was numerically obtained from the imaginary part of the diagonal components of the self-energy. In the pristine model, this quantity exhibits a characteristic sixfold angular symmetry in $\textbf{q}$ and a linear scaling with temperature~\cite{ye2020phonon}. In the disordered system, we found that a sixfold pattern is also present, while the temperature evolution strongly depends on the value of $\kappa$ and the sound velocity $v_s$. Through power-counting estimation, we demonstrated that the generic form of the temperature evolution must depend on the density of states and the Fermi-Dirac distribution for each scattering channel.

In our full numerical analysis, we observed an approximate linear scaling with temperature in the experimentally accessible range for $\kappa = 0$, regardless of the quasivacancy concentration. However, when $\kappa \neq 0$, the results show a distinct linear scaling with temperature only if we fine-tune the phonon frequency $\Omega$ while keeping the sound velocity $v_s$ constant. This fine-tuning is necessary due to the localized nature of the in-gap states induced by the quasivacancies, which require a specific value of incoming phonon energy to scatter into the bulk, thus contributing to the ph-channel. This effect is absent in the clean limit, where the Majoranas are fully gapped when $\kappa \neq 0$.

Finally, we explored the temperature scaling of $\alpha_s(\textbf{q})$ in the random-flux sector, as a limiting case to capture the physics of thermally excited fluxes at temperatures above the flux gap scale. Using a power-counting argument, we demonstrated that although flux disorder overrides the effects of quasivacancies, the attenuation still scales with temperature as $\alpha_s \sim \alpha_0 + \alpha_1 T$. This is because ph-processes grow linearly with temperature within a smaller window in the random-flux sector. Combined with dominant pp-processes, this results in a bounded scaling with temperature.
Our numerical analysis corroborates this observation, highlighting the robustness of sound attenuation in the presence of both site and flux disorder. This robustness is due to the fermionic nature of fractionalized excitations, in contrast to other bosonic contributions such as phonon-phonon and magnon-phonon scatterings. In conclusion, our combined results reinforce the effectiveness of acoustic probes as a means to experimentally verify spin fractionalization in Kitaev materials.

\section{Acknowledgements}
The authors thank  Susmita Singh, Peter Stavropoulos, Swetlana Swarup and Yang Yang for valuable discussions. The work is supported by the U.S. Department of Energy, Office of Science, Basic Energy Sciences under Award No. DE-SC0018056.   We acknowledge the support from NSF DMR-2310318 and the support of the Minnesota Supercomputing Institute (MSI) at the University of Minnesota. N.B.P. also acknowledges the hospitality and partial support of the Technical University of Munich – Institute for Advanced Study and the Alexander von Humboldt Foundation.

%
%
%
\appendix


\section{Computation of the phonon polarization bubble} \label{App: calc_details}

 In this appendix, we provide some technical details for the computation of the phonon polarization bubble.
\subsection{Majorana fermion-phonon vertex coupling in the mixed representation}\label{App:MFP_vertex}

As shown in the main text, the matrix elements of the Majorana fermion-phonon vertex in the mixed representation are given by 
\begin{align}
	\lamb{ij} = 2i\lambda \eta^\alpha_{ij} f^\mu_\alpha(\textbf{q}),
\end{align}
where the functions $f^\mu_\alpha(\q)$ are obtained from the Fourier transform of  the strain
tensor $\epsilon_{ij}(\textbf{r})$ in the basis of the phonon polarization vectors. The explicit form of $f^\mu_\alpha(\q)$ is:
\begin{align}
	&\begin{cases}
		&f_x^\parallel(\textbf{q}) = \pa{q_x + \sqrt{3}q_y}\cos\theta_\textbf{q}+ \pa{\sqrt{3}q_x - q_y} \sin\theta_\textbf{q},  \\
		&f_y^\parallel(\textbf{q}) = \pa{q_x - \sqrt{3}q_y}\cos\theta_\textbf{q}- \pa{\sqrt{3}q_x + q_y} \sin\theta_\textbf{q},  \\
		&f_z^\parallel(\textbf{q}) = 2\pa{-q_x\cos\theta_\textbf{q} +q_y\sin\theta_\textbf{q} }, 
	\end{cases}\\
	&	
	\begin{cases}
		&f_x^\perp(\textbf{q}) =\pa{\sqrt{3}q_x - q_y}\cos\theta_\textbf{q}- \pa{q_x + \sqrt{3}q_y} \sin\theta_\textbf{q},  \\
		&f_y^\perp(\textbf{q}) =-\pa{\sqrt{3}q_x + q_y}\cos\theta_\textbf{q}- \pa{q_x - \sqrt{3}q_y} \sin\theta_\textbf{q}, \\
		&f_z^\perp(\textbf{q}) =  2\pa{q_x\sin\theta_\textbf{q} +q_y\cos\theta_\textbf{q} }. 
	\end{cases}
\end{align}

The next step is to absorb the exponential factors from  Eq.~(\ref{eq:Vqmajs}) into our definition of $\lamb{ij}$. To achieve this, we define the exponential matrices   $\mathbb{E}_\q^{{A,B}}$, which contain all the phase factors corresponding to sites on each sublattice: 
\begin{align}
	&\mathbb{E}_\q^{\text{A}} = 
	\begin{pmatrix}
		e^{i\q \cdot \textbf{r}_1} & \dots & e^{i\q \cdot \textbf{r}_1} & \dots & e^{i\q \cdot \textbf{r}_1}\\
		\dots & \dots  & \dots & \dots  & \dots  \\
		e^{i\q \cdot \textbf{r}_i} & \dots & e^{i\q \cdot \textbf{r}_i} & \dots & e^{i\q \cdot \textbf{r}_i}\\
		\dots & \dots  & \dots & \dots  & \dots  \\
		e^{i\q \cdot \textbf{r}_N} & \dots & e^{i\q \cdot \textbf{r}_N} & \dots & e^{i\q \cdot \textbf{r}_N}\\
	\end{pmatrix}, \\ 
	&\co{\mathbb{E}_\q^{\text{B}}}^T = 
	\begin{pmatrix}
		& e^{i\q \cdot \textbf{r}_{1}} & \vdots & e^{i\q \cdot \textbf{r}_j} & \vdots & e^{i\q \cdot \textbf{r}_N} \\
		& \vdots & \vdots  & \vdots & \vdots  & \vdots  \\
		& e^{i\q \cdot \textbf{r}_1} & \vdots & e^{i\q \cdot \textbf{r}_j} & \vdots & e^{i\q \cdot \textbf{r}_N} \\
		& \vdots & \vdots & \vdots\ & \vdots\ & \vdots \\
		& e^{i\q \cdot \textbf{r}_1} & \vdots & e^{i\q \cdot \textbf{r}_j} & \vdots & e^{i\q \cdot \textbf{r}_N}. 
	\end{pmatrix}
\end{align}
Here all $\textbf{r}_i$ inside $\mathbb{E}_\q^A$ belongs to sublattice A, and all $\textbf{r}_i$ in $\mathbb{E}_\q^B$ belongs to B.  
This allows us to write the full vertex matrix $\Lambda_{\q}^\mu$ by taking the element-wise (Hadamard) product of the vertex $\lamb{ij}$ with the exponential matrices:
\begin{align}
	\Lambda_{\q}^\mu = 	\begin{pmatrix}
		0 & \lambda_{\q}^\mu \\
		-\co{\lambda_{\q}^\mu}^T& 0 
	\end{pmatrix} \odot
	\begin{pmatrix}
		0 & \mathbb{E}^A_\q \\
		\co{\mathbb{E}^B_\q}^T& 0 
	\end{pmatrix}.
\end{align}
Here we used the fact that the matrix from sublattice $B$ to $A$ is simply:
\begin{align}
	\lamb{BA} = \co{\lamb{AB}}^\dagger = -\co{\lamb{AB}}^T, \label{lambABtoBA}
\end{align}
where the last step comes from the fact that $\lamb{ij}$ is purely imaginary.  This allows  us to obtain  the coupling $V_\q$   defined in Eq.~(\ref{eq:Vqmajs})  in the sublattice representation and recover Eq.~(\ref{eq:Vq_matrix}):
\begin{align}\label{eq:Vq_matrix_app}
	V_\q = -\frac{i}{2}
	\begin{pmatrix}
		c_A & c_B
	\end{pmatrix}
	\begin{pmatrix}
		0 & \Lamb{AB}\\
		\Lamb{BA}& 0
	\end{pmatrix}
	\begin{pmatrix}
		c_A\\c_B
	\end{pmatrix}\widetilde{u}_\q^\mu
\end{align}
The submatrices $\Lamb{AB}$ and $\Lamb{BA}$ are then given by:
\begin{align}
\begin{split}
	&\Lamb{AB} = \lambda_\q^{\mu} \odot \mathbb{E}_\q^A, \\
	&\Lamb{BA} = -\co{\lambda_\q^\mu}^T \odot \co{\mathbb{E}_\q^B}^T  .
\end{split}
\end{align}
As shown in \cite{feng2021temperature}, it is useful to further simplify this expression by symmetrizing the coupling matrix between between A and B sublattices. The elements of the symmetrized exponential matrices are then:
\begin{align}
	[\mathbb{E}^s_{\q}]_{ij} = \frac{1}{2}\pa{e^{i\q\cdot\textbf{r}_i} + e^{i\q\cdot\textbf{r}_j}}.
\end{align}
This symmetrization ensures that the coupling matrix accounts for the interactions between the two sublattices in a balanced manner and 
 recover  the expressions for submatrices $\Lamb{AB}$ and $\Lamb{BA}$ presented in
 Eq.~\eqref{Lamb_matelem}. The final form of the symmetrized vertex is the same as in Eq.~\eqref{eq:Vq_matrix_app}, but with the symmetrized submatrices instead. From this point, we can drop the superscript $s$, as all the sublattice vertex matrices are assumed to take the form presented in the main text.

	In principle, we could proceed with the self-energy calculation from the expression above. However, it is instructive to see how each block is written explicitly in terms of the Bogoliubov submatrices $X$ and $Y$. By taking the matrix product explicitly in 
 Eq.~\eqref{lambtildef} we have

\begin{widetext}
\begin{align}
\begin{split}
		&\widetilde{\Lambda}^\mu_{\q,11} = i\co{ \pa{X-Y}^*\Lamb{BA}\pa{X+Y}^T - \pa{X+Y}^*\Lamb{AB}\pa{X-Y}^T }\\
		&\widetilde{\Lambda}^\mu_{\q,12} = i\co{ \pa{X-Y}^*\Lamb{BA}\pa{X+Y}^\dagger + \pa{X+Y}^*\Lamb{AB}\pa{X-Y}^\dagger }\\
		&\widetilde{\Lambda}^\mu_{\q, 21} = -i\co{ \pa{X-Y}\Lamb{BA}\pa{X+Y}^T + \pa{X+Y}\Lamb{AB}\pa{X-Y}^T }\\
		&\widetilde{\Lambda}^\mu_{\q,22} = -i\co{ \pa{X-Y}\Lamb{BA}\pa{X+Y}^\dagger - \pa{X+Y}\Lamb{AB}\pa{X-Y}^\dagger }
\end{split}
\end{align}
where $\Lamb{BA}$ can be written in terms of $\Lamb{AB}$ by using the property of the symmetrized vertex in 
Eq.~\eqref{eq:lambda_BAtoAB}. 
The advantage of using the above expression is that $X$ and $Y$ are $N\times N$ matrices such that their matrix multiplication is practically much more efficient than that of the  $2N\times 2N$ matrices, $W$ and $U$, defined in Eq.~\eqref{lambtildef}.
	
	\subsection{Explicit steps to obtain $\Pi^{\mu\nu}(\textbf{q},\tau)$}\label{App:steps_PI}
	
Here we show some extra steps to obtain the phonon polarization bubble shown in Sec.~\ref{subsec:phonon_pol_bubble}. From Eq.~\eqref{Vqbogo}, we can take the matrix product to write Eq.~\eqref{Pi_def} in a more explicit form:
	\begin{align}
		\Pi^{\mu\nu}(\q,\tau) =& \left\langle T_\tau 
		\co{ \beta^\dagger \tLambij{\mu}{11} \beta + \beta^\dagger \tLambij{\mu}{12} \beta^\dagger + \beta \tLambij{\mu}{21} \beta + \beta \tLambij{\mu}{22} \beta^\dagger }(\tau) \times \right.\\
		&\hspace{6mm} \left.\co{ \beta^\dagger \tLambmij{\nu}{11} \beta + \beta^\dagger \tLambmij{\nu}{12} \beta^\dagger + \beta \tLambmij{\nu}{21} \beta + \beta \tLambmij{\nu}{22} \beta^\dagger }(0) \right\rangle .\nonumber
	\end{align}
By using Wick's theorem, one can show that there are only six non-vanishing contributions out of the sixteen terms in the expression above. For instance, one possible term in the 
ph-channel can be written as:
\begin{align}
\begin{split}\label{eqapp_wick}
		\expval{T_\tau  \co{ \beta^\dagger \tLambij{\mu}{11} \beta}(\tau) \co{ \beta^\dagger \tLambmij{\nu}{11} \beta}(0)} 
		&= \expval{T_\tau  \co{ \beta^\dagger_i \beta_j}(\tau)\co{ \beta^\dagger_k  \beta_l}(0)}
		\co{\tLambij{\mu}{11} }_{ij}\co{\tLambmij{\nu}{11}}_{kl}  \\
		&= \expval{T_\tau \beta^\dagger_i(\tau)\beta_l(0)}\expval{T_\tau  \beta_j(\tau)\beta_k^\dagger(0)}\co{\tLambij{\mu}{11} }_{ij}\co{\tLambmij{\nu}{11}}_{kl} \\
		  & = \co{\gb_i(\tau)g_j(\tau)}\delta_{il}\delta_{jk}\co{\tLambij{\mu}{11} }_{ij}\co{\tLambmij{\nu}{11}}_{kl} \\
		& = \co{\gb_i(\tau)g_j(\tau)}\co{[\tLambij{\mu}{11}] \odot [\tLambmij{\nu}{11}]^T}_{ij},
\end{split}
\end{align}
where $g_i(\tau) = \mean{T_\tau \beta_i(\tau)\beta^\dagger_i(0)}$ and $\bar{g}_i(\tau) = \mean{T_\tau \beta^\dagger_i(\tau)\beta_i(0)}$ are the fermionic propagators.  The summation over repeated indices is assumed in \eqref{eqapp_wick}, and the definition of the Hadamard product was used. The same kind of computation can be easily performed for all non-vanishing terms. From this, we can sum all contributions to write the polarization bubble in terms of the imaginary time:
	
 \begin{align}\label{eq:pi_taufull}
 \begin{split}
		\Pi^{\mu\nu}(\q,\tau) = \frac{1}{N}\sum_{ij} &\co{\gb_i(\tau)g_j(\tau)}\co{[\tLambij{\mu}{11}]\odot[\tLambmij{\nu}{11}]^T -[\tLambij{\mu}{11}]\odot[\tLambmij{\nu}{22}] }_{ij}\\
		&\hspace{-3.5mm}+\co{g_i(\tau)\gb_j(\tau)}\co{[\tLambij{\mu}{22}]\odot[\tLambmij{\nu}{22}]^T -[\tLambij{\mu}{22}]\odot [\tLambmij{\nu}{11}] }_{ij} \\
		&\hspace{-3.5mm}+\co{g_i(\tau)g_j(\tau)}\co{[\tLambij{\mu}{21}]\odot [\tLambmij{\nu}{12}]^T -[\tLambij{\mu}{21}]\odot [\tLambmij{\nu}{12}] }_{ij} \\
        &\hspace{-3.5mm}+\co{\gb_i(\tau)\gb_j(\tau)}\co{[\tLambij{\mu}{12}]\odot[\tLambmij{\nu}{21}]^T -[\tLambij{\mu}{12}]\odot [\tLambmij{\nu}{21}] }_{ij}.
\end{split}	
\end{align}
\end{widetext}
The final step is to take the Fourier transform to the Matsubara frequencies $i\omega_n$.

The fermionic Matsubara Green's functions are given by:
\begin{equation}
	g_i(i\omega_n) = \frac{1}{i\omega_n - \varepsilon_i}\, , \qquad \bar{g}_i(i\omega_n) = \frac{1}{i\omega_n + \varepsilon_i},
\end{equation}
which are obtained from $g_i(i\omega_n) = \int_0^\beta\, d\tau e^{i\omega_n\tau} g_i(\tau)$, and the same for $\bar{g}_i(i\omega_n)$.
Using this form, along with the Fourier transform over the phonon frequencies, we can compute the convolutions of the Green's functions in Eq.~\eqref{eq:piomega_full} as:
\begin{align}
\begin{split}
	P_{ij}^{g\bar{g}} &= T\sum_{\omega_n} g_i(i\omega_n)\bar{g}_j(i\Omega - i\omega_n)\\
	& = T\sum_{\omega_n}\frac{1}{i\omega_n - \varepsilon_i}\frac{1}{(i\Omega - i\omega_n) + \varepsilon_j}\\
	& = \frac{n_F(\varepsilon_i) - n_F(\varepsilon_j)}{i\Omega - \varepsilon_i + \varepsilon_j},
\end{split}
\end{align}
\begin{align}
\begin{split}
	P_{ij}^{\bar{g}g} &= T\sum_{\omega_n} \bar{g}_i(i\omega_n)g_j(i\Omega - i\omega_n)\\
	& = T\sum_{\omega_n}\frac{1}{i\omega_n + \varepsilon_i}\frac{1}{(i\Omega - i\omega_n) - \varepsilon_j}\\
	& = \frac{n_F(-\varepsilon_i) - n_F(-\varepsilon_j)}{i\Omega + \varepsilon_i - \varepsilon_j},
\end{split}
\end{align}
\begin{align}
\begin{split}
	P_{ij}^{gg} &= T\sum_{\omega_n} g_i(i\omega_n)g_j(i\Omega - i\omega_n)\\
	& = T\sum_{\omega_n}\frac{1}{i\omega_n - \varepsilon_i}\frac{1}{(i\Omega - i\omega_n) - \varepsilon_j}\\
	& = \frac{n_F(\varepsilon_i) - n_F(-\varepsilon_j)}{i\Omega - \varepsilon_i - \varepsilon_j},
\end{split}
\end{align}
\begin{align}
\begin{split}
	P_{ij}^{\bar{g}\bar{g}} &= T\sum_{\omega_n} \bar{g}_i(i\omega_n)\bar{g}_j(i\Omega - i\omega_n)\\
	& = T\sum_{\omega_n}\frac{1}{i\omega_n + \varepsilon_i}\frac{1}{(i\Omega - i\omega_n) + \varepsilon_j}\\
	& = \frac{n_F(-\varepsilon_i) - n_F(\varepsilon_j)}{i\Omega + \varepsilon_i + \varepsilon_j},
\end{split}
\end{align}
where we used the method of residues to compute the sum over complex frequencies \cite{mahan2000many}. Therefore, we can recover Eq.~(\ref{Pggbar}) given in the main text by taking an analytical continuation $\Omega \rightarrow \Omega + i\delta$ for infinitesimal $\delta$.

\begin{figure*}
	\centering
	\includegraphics[width=0.85\linewidth]{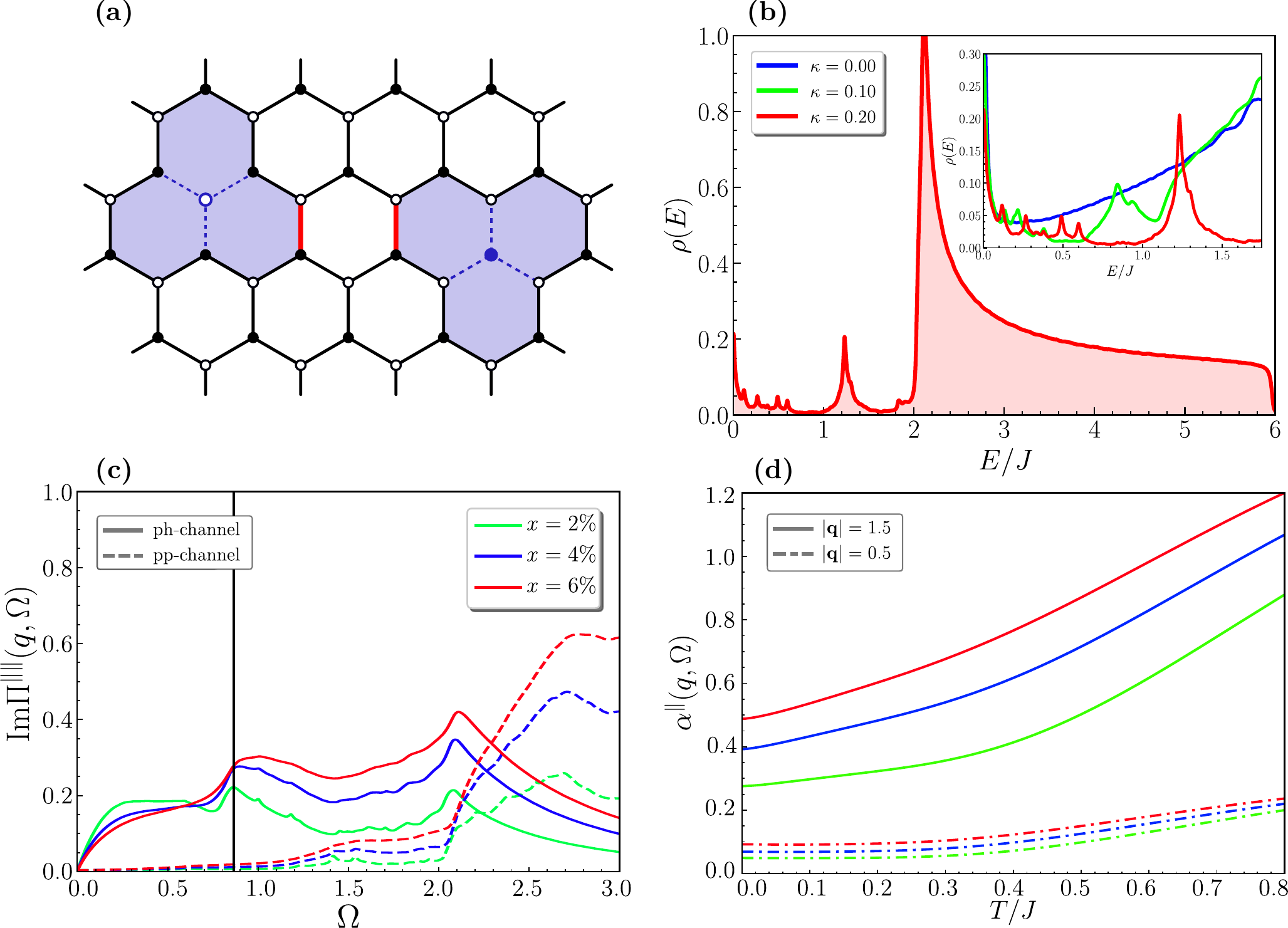}
	\caption{(a) A pair of quasivacancies in the bound-flux sector. The removed sites on each sublattice are marked in blue, and the weak couplings $J^\prime$ are denoted by the dashed line. Red bonds indicate the negative link variables and the $\pi$-fluxes on the vacancy plaquette are shaded. (b) The density of states in the bound-flux sector with $\kappa = 0.2$ and  $x = 2\%$. Inset: In-gap peak for different values of $\kappa$. (c)  The imaginary part of the phonon self-energy. The contribution from each channel is plotted as a function of the $\Omega$ with $|\textbf{q}| = 0.5$, $\theta_{\mathbf{q}} = \pi/6$, $T = 0.8J$, and different values of $x$. The vertical line indicates the energy $\Omega = v_F q$. (d) $\alpha^{\parallel}(\textbf{q})$ as a function of temperature for different values of $x$ and two values of $|\textbf{q}|$, both with $\theta_\textbf{q} = \pi/6$. All the results are averaged over 1000 disorder realizations with $L=32$, and $J^\prime = 0.01 J$.}
	\label{fig:app}
\end{figure*}
\section{Results for the bound-flux sector}\label{App:bound-flux}

In this appendix, we show some results for the sound attenuation in the bound-flux sector, with one possible realization is shown in Fig.~\ref{fig:app}(a). As outlined in previous works \cite{kao2021vacancy,dantas2022}, a clear distinction between the bound- and zero-flux sectors emerges for $\kappa \neq 0$. Such difference is evident from the density of states, where in contrast to the zero-flux one additional resonant peak is present around $E = 0$ in the bound-flux sector, independent of the strength of $\kappa$ (Fig.~\ref{fig:app}(b)). 

The appearance of such localized mode could, in principle, change the response of the system in the attenuation process for $\kappa\neq 0$, which is evidenced in the imaginary part of the self-energy as a function of the phonon frequency $\Omega$, as plotted in Fig.~\ref{fig:app}(c). 
Because of the additional in-gap states in the bound-flux sector, the overlap of the low-frequency peaks leads to a larger range of $\Omega$ where the ph-channel is appreciable.
This behavior is in visible contrast to the zero-flux sector.
These new features are, however, almost insensible to the temperature scaling of the sound attenuation. As shown in Fig.~\ref{fig:app}(d), there is still a threshold of values of $|\q|$ where the linear scaling in temperature is observable, for values close to the ones in the zero-flux sector. Again, this is a direct consequence of the new manifold of low-energy states accessible in the 
ph-scattering process. Since we do not expect such fine-tuning to be achievable in ultrasound experiments, we conclude that the bound-flux and zero-flux sectors are nearly indistinguishable from the perspective of acoustic phonon dynamics.



%
%

\bibliographystyle{apsrev4-1} 
\bibliography{refs} 

\end{document}